\documentclass[a4paper,11pt]{article}
\usepackage{preamble/jheppubMOD} % for details on the use of the package, please
\usepackage{lipsum}
\usepackage{graphicx}  % needed for figures
\usepackage{multirow}
\usepackage[x11names,dvipsnames]{xcolor}

\usepackage{preamble/mciteplus}

\hypersetup{
    colorlinks,
    linkcolor={jpac-blue},
    citecolor={jpac-blue},
    urlcolor={jpac-blue}
}

\usepackage{lmodern}
\usepackage{amsmath,amssymb}
\usepackage{mathtools}
\usepackage[numbers,sort&compress]{natbib}

\usepackage{tikz}
\usepackage{tikz-3dplot}
\usepackage{pgfplots,pgfplotstable}
\usetikzlibrary{pgfplots.groupplots}
\usetikzlibrary{decorations.markings}
\usetikzlibrary{shapes}
\usetikzlibrary{shapes.misc}
\usetikzlibrary{patterns}
\usepgfplotslibrary{fillbetween}
\pgfplotsset{compat=newest}
\usetikzlibrary{positioning}
\usetikzlibrary{calc}

\usepackage{lipsum}
\usepackage{stackrel}

\usepackage[compat=1.1.0]{tikz-feynman}

\usepackage{cleveref}
\usepackage{ifthen}
\usepackage{xspace}
\usepackage[shortlabels]{enumitem}

\definecolor{jpac-blue}{RGB}{ 31,119,180}%  hex = #1F77B4
\definecolor{jpac-red}{RGB}{214,39, 40}%  hex = #D61D28
\definecolor{jpac-green}{RGB}{ 44,160, 44}%  hex = #2CA02C
\definecolor{jpac-orange}{RGB}{255,127, 14}%  hex = #FF7F0E
\definecolor{jpac-purple}{RGB}{148,103,189}%  hex = #9467BD
\definecolor{jpac-brown}{RGB}{140, 86, 75}%  hex = #8C564B
\definecolor{jpac-pink}{RGB}{227,119,194}%  hex = #E377C2
\definecolor{jpac-gold}{RGB}{188,189, 34}%  hex = #BCBD22
\definecolor{jpac-aqua}{RGB}{ 23,190,207}%  hex = #17BECF
\definecolor{jpac-grey}{RGB}{127,127,127}%  hex = #7F7F7F

\pgfdeclarepatternformonly{crosshatchD}{\pgfqpoint{-1.5pt}{-1.5pt}}{\pgfqpoint{4pt}{4pt}}{\pgfqpoint{2.5pt}{2.5pt}}%
{%
  \pgfsetlinewidth{0.4pt}%
  \pgfpathmoveto{\pgfqpoint{3.1pt}{0pt}}%
  \pgfpathlineto{\pgfqpoint{0pt}{3.1pt}}%
  \pgfpathmoveto{\pgfqpoint{0pt}{0pt}}%
  \pgfpathlineto{\pgfqpoint{3.1pt}{3.1pt}}%
  \pgfusepath{stroke}%
}%

\tikzset{vpiFF/.style={draw=jpac-red,fill=jpac-pink!40 ,circle,very thick,inner sep=4.5pt}}
\tikzset{vfone/.style={draw=jpac-blue,fill=jpac-aqua!40,circle,very thick,inner sep=4.5pt}}

\tikzfeynmanset{pion/.style={black,charged scalar, dash pattern=on 3pt off 2pt,very thick}}
\tikzfeynmanset{phot/.style={black,photon,very thick}}
\tikzfeynmanset{omeg/.style={black,fermion,very thick}}
\tikzfeynmanset{cutAs/.style={black!80,very thick}}
\tikzfeynmanset{cutBs/.style={black!80,thin,decoration={pre length=0.06cm, post length=0.0cm,border,segment length=0.4mm,amplitude=1mm,angle=270},
            postaction={decorate,draw}}}

\pgfplotsset{error bar legend/.style={%
    /pgfplots/legend image code/.prefix code={%
      \pgfkeysgetvalue{/pgfplots/error bars/error mark}{\pgfplotserrorbarsmark}%
      \draw[%
        /pgfplots/every error bar, 
        mark=\pgfplotserrorbarsmark, 
        /pgfplots/error bars/error mark options, 
        sharp plot,
        ##1,
      ] plot coordinates {(0.30cm, -0.15cm) (0.30cm, 0.15cm)};%
    }
  }
}

\pgfplotsset{DataNA6009/.style={line width=1pt,color=jpac-orange,only marks, mark=*,mark size=1.35pt,opacity=0.65, fill opacity=0.75,on layer={axis background},error bar legend,error bars/.cd, y dir=both,y explicit,error bar style={opacity=0.75,fill opacity=0.75,line width=0.75pt}}}

\pgfplotsset{DataNA6016/.style={line width=1pt,color=jpac-green,only marks, mark=triangle*,mark size=1.35pt,opacity=0.65, fill opacity=0.75,on layer={axis background},error bar legend,error bars/.cd, y dir=both,y explicit,error bar style={opacity=0.75,fill opacity=0.75,line width=0.75pt}}}

\pgfplotsset{DataMAMI/.style={line width=1pt,color=jpac-purple,only marks, mark=square*,mark size=1.35pt,opacity=0.65, fill opacity=0.75,on layer={axis background},error bar legend,error bars/.cd, y dir=both,y explicit,error bar style={opacity=0.75,fill opacity=0.75,line width=0.75pt}}}

\colorlet{lowphi}{jpac-blue}
\colorlet{higphi}{jpac-red}
\colorlet{expcolor}{jpac-green}

\pgfplotsset{HistoStyleL/.style={const plot,draw=none,fill=lowphi,opacity=0.5}}
\pgfplotsset{HistoStyleH/.style={const plot,draw=none,fill=higphi,opacity=0.5}}

\newcommand\bsub{\begin{subequations}}
\newcommand\esub{\end{subequations}}

\newcommand{\eg}{{\it e.g.}\xspace}
\newcommand{\ie}{{\it i.e.}\xspace}
\newcommand{\cf}{{\it cf.}\xspace}

\newcommand{\kev}{\ensuremath{{\mathrm{\,ke\kern -0.1em V}}}\xspace}
\newcommand{\mev}{\ensuremath{{\mathrm{\,Me\kern -0.1em V}}}\xspace}
\newcommand{\gev}{\ensuremath{{\mathrm{\,Ge\kern -0.1em V}}}\xspace}
\newcommand{\gevsq}{\ensuremath{{\mathrm{\,Ge\kern -0.1em V}^2}}\xspace}
\newcommand{\tev}{\ensuremath{{\mathrm{\,Te\kern -0.1em V}}}\xspace}

\def\XXint#1#2#3{{\setbox0=\hbox{$#1{#2#3}{\int}$}\vcenter{\hbox{$#2#3$}}\kern-.5\wd0}}

\newcommand{\moda}{\left\lvert a \right\rvert}
\newcommand{\modb}{\left\lvert b \right\rvert}
\newcommand{\phab}{\phi_b}
\newcommand{\modf}{\left\lvert f_{\omega\pi^0}(0) \right\rvert}
\newcommand{\phaf}{\phi_{\omega\pi^0}(0)}
\newcommand{\fwsq}{\left\lvert F_{\omega\pi}(s) \right\rvert^2}

\usepgfplotslibrary{external}
\usetikzlibrary{external}
\pgfrealjobname{omegato3pi}

\begin{document}

\title{\boldmath $\omega \to 3\pi$ and $\omega\pi^{0}$ transition form factor revisited}

\collaboration{JPAC Collaboration}
\collaborationImg{\includegraphics[height=2cm,keepaspectratio]{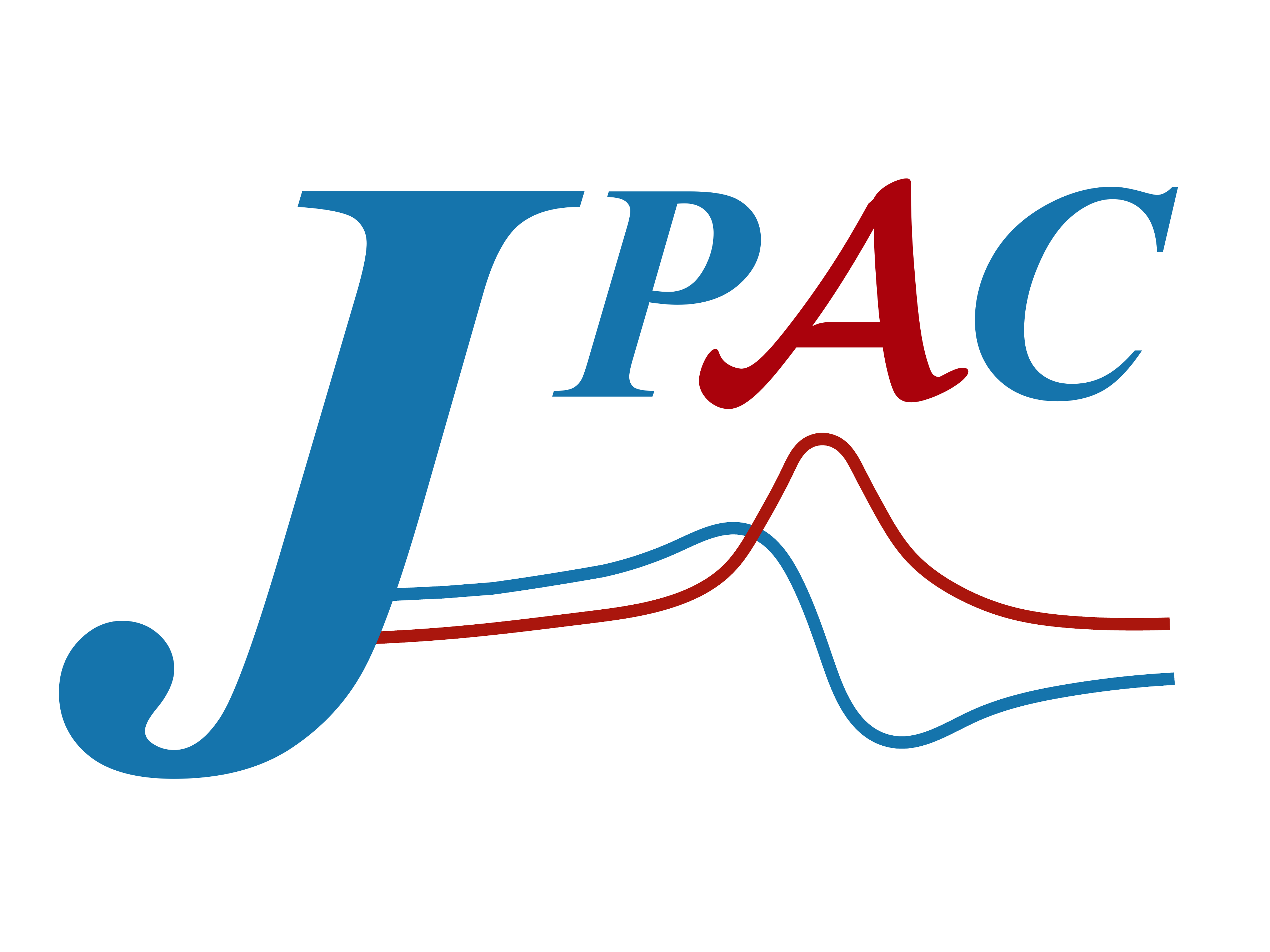}}
\preprint{JLAB-THY-20-3200}

%%Affiliations
\newcommand{\ceem}{Center for  Exploration  of  Energy  and  Matter,
Indiana  University,
Bloomington,  IN  47403,  USA}
\newcommand{\indiana}{Department of Physics,
Indiana  University,
Bloomington,  IN  47405,  USA}
\newcommand{\jlab}{Theory Center,
Thomas  Jefferson  National  Accelerator  Facility,
Newport  News,  VA  23606,  USA}
\newcommand{\icn}{Instituto de Ciencias Nucleares, 
Universidad Nacional Aut\'onoma de M\'exico, Ciudad de M\'exico 04510, Mexico}
\newcommand{\hiskp}{Universit\"at Bonn,
Helmholtz-Institut f\"ur Strahlen- und Kernphysik, 53115 Bonn, Germany}
\newcommand{\ect}{European Centre for Theoretical Studies in Nuclear Physics and related Areas (ECT$^*$) and Fondazione Bruno Kessler, Villazzano (Trento), I-38123, Italy}
\newcommand{\genova}{INFN Sezione di Genova, Genova, I-16146, Italy}
\newcommand{\cern}{CERN, 1211 Geneva 23, Switzerland}
\newcommand{\ucm}{Departamento de F\'isica Te\'orica, Universidad Complutense de Madrid and IPARCOS, 28040 Madrid, Spain}
\newcommand{\mainz}{Institut f\"ur Kernphysik \& PRISMA$^+$  Cluster of Excellence, Johannes Gutenberg Universit\"at,  D-55099 Mainz, Germany}

\author[a]{M.~Albaladejo}
\emailAdd{albalade@jlab.org}
\author[b]{I.~Danilkin}
\emailAdd{danilkin@uni-mainz.de}
\author[c,d]{S.~Gonz\`{a}lez-Sol\'is}
\emailAdd{sgonzal@iu.edu}
\author[c,d]{D.~Winney}
\emailAdd{dwinney@iu.edu}
\author[e] {C.Fern\'andez-Ram\'irez}
\author[a]{A.~N.~Hiller Blin}
\author[f]{V.~Mathieu}
\author[g]{M.~Mikhasenko}
\author[h,i]{A.~Pilloni}
\author[a,c,d]{A.~Szczepaniak}

\affiliation[a]{\jlab}
\affiliation[b]{\mainz}
\affiliation[c]{\indiana}
\affiliation[d]{\ceem}
\affiliation[e]{\icn}
\affiliation[f]{\ucm}
\affiliation[g]{\cern}
\affiliation[h]{\ect}
\affiliation[i]{\genova}

\abstract{In light of recent experimental results, we revisit the dispersive analysis of the $\omega \to 3\pi$ decay amplitude and of the $\omega\pi^0$ transition form factor. Within the framework of the Khuri-Treiman equations, we show that the $\omega \to 3\pi$ Dalitz-plot parameters obtained with a once-subtracted amplitude are in agreement with the latest experimental determination by BESIII. Furthermore, we show that at low energies the $\omega\pi^0$ transition form factor obtained from our determination of the $\omega \to 3\pi$ amplitude is consistent with the data from MAMI and NA60 experiments.}

\frenchspacing
%\toccontinuoustrue
\maketitle
\newpage

%***********************************************
%% Main Body
\section{Introduction}\label{sec:introduction}

A precise description of the amplitudes involving three particles in the final state is one of the open challenges in hadron physics. It becomes even more important in view of the high precision data from the GlueX, CLAS12, COMPASS, BESIII, and LHCb experiments, where various exotic states decaying to three particles have been or will be measured \cite{Aaij:2014jqa,Aaij:2015tga,Ghoul:2015ifw,Krinner:2017vch,Akhunzyanov:2018pnr,Ablikim:2019hff,Mokeev:2020hhu}. A proper description of three-particle amplitudes is also required for extraction of resonance parameters from lattice QCD computations~\cite{Briceno:2017max,Jackura:2019bmu,Briceno:2019muc,Mai:2019pqr,Culver:2019vvu}.

At low energies the adequate formalism to treat the three-body decays is based on the so-called Khuri-Treiman (KT) equations~\cite{Khuri:1960zz} which make the maximal use of analyticity, unitarity, and crossing symmetry via dispersion relations. They were extensively applied in the study of the isospin breaking $\eta\rightarrow 3\pi$ decay~\cite{Guo:2014vya,Guo:2015zqa,Guo:2016wsi,Colangelo:2016jmc,Colangelo:2018jxw, Albaladejo:2017hhj, Gasser:2018qtg}, and several other reactions~\cite{Niecknig:2012sj, Danilkin:2014cra, Niecknig:2015ija, Isken:2017dkw, Niecknig:2017ylb}, and later generalized to include arbitrary spin, isospin, parity, and charge conjugation for the decaying particle~\cite{Albaladejo:2019huw} (see also Ref.~\cite{Mikhasenko:2019rjf}). Among the various applications, the decay of light vector isoscalar resonances $\omega/\phi \to 3\pi$ serves as one of the benchmark cases for  dispersion theory. Because of Bose symmetry only odd angular momentum is allowed in each of the $\pi\pi$ channels, and thus the final state is dominated by the $J=I=1$ isobars, \ie the $\rho$ meson. The latter is related to the $J=I=1$ $\pi\pi$ partial wave amplitude, which is known to high precision from the Roy analysis of $\pi\pi$ scattering~\cite{GarciaMartin:2011cn}. The existing analyses of the decays of $\omega$ and $\phi$ to three pions~\cite{Niecknig:2012sj,Danilkin:2014cra,Dax:2018rvs} are mainly based on unsubtracted dispersion relations, which result in a parameter-free prediction of the shape of the Dalitz plot. The $\phi$ decay was favorably compared to the high-statistics Dalitz-plot data from the KLOE~\cite{Aloisio:2003ur} and CMD-2~\cite{Akhmetshin:2006sc} experiments. Until recently the only available data for $\omega \to 3\pi$ came from the WASA-at-COSY experiment~\cite{Adlarson:2016wkw}. Given that the nominal $\rho\pi$ threshold is above the mass of the $\omega$, the distribution of events in the Dalitz plot is rather smooth and, therefore, it can be efficiently parametrized in the experimental analyses by a low-order polynomial in the Dalitz-plot variables. The coefficient of the leading term in the Dalitz-plot polynomial expansion (\ie the Dalitz-plot parameter $\alpha$) obtained this way is consistent with the dominance of the $\rho$ peak, even though it lies outside the kinematical boundary. However, the experimental uncertainties in the WASA-at-COSY measurement were too large to verify the prediction of dispersion relation calculations.

The situation changed recently when the high-statistics data from BESIII became available~\cite{Ablikim:2018yen}. A new set of $\omega \to 3\pi$ Dalitz-plot parameters was extracted from the data and was found to differ significantly from the predictions based on (unsubtracted) dispersion relations of the KT equations~\cite{Niecknig:2012sj,Danilkin:2014cra}. This is particularly unsettling since, as mentioned above, there is good agreement between the data and the predicted shape of the Dalitz plot in the case of the $\phi$ decays~\cite{Niecknig:2012sj}. Therefore, in this paper we reanalyze the $\omega \to 3\pi$ BESIII data with the KT equations. 

As in any low-energy effective theory, the contribution from inelastic channels enters as (free) parameters, and this can be the origin of the discrepancy between the data and the calculation based on the unsubtracted dispersion relations. We also reanalyze the $\omega\pi^0$ transition form factor (TFF), which controls the $\omega \to \pi^0 \gamma^{(\ast)}$ amplitude. At low energies, the TFF is sensitive to the $\omega \to 3\pi$ amplitude. There are recent data on the form factor from the MAMI~\cite{Adlarson:2016hpp} and NA60~\cite{Arnaldi:2009aa,Arnaldi:2016pzu} collaborations. As we will see below, the simultaneous analysis of both reactions allows one to better constrain the subtraction constant of the $\omega \to 3\pi$ amplitude.

The analysis presented in this paper could also be relevant to understand the hadronic contributions to the anomalous magnetic moment of the muon. The presently observed $\sim\!3\sigma$ deviation between theory~\cite{Jegerlehner:2017gek, Keshavarzi:2018mgv, Davier:2017zfy, Danilkin:2019mhd} and experiment~\cite{Tanabashi:2018oca} has a potential to become more significant once new measurements at both Fermilab~\cite{LeeRoberts:2011zz, Grange:2015fou} and J-PARC~\cite{Iinuma:2011zz} become available. The theoretical uncertainties mainly originate from the hadronic vacuum polarization (HVP) and the hadronic light-by-light (HLbL) processes, with $\omega \to 3\pi, \pi^0 \gamma^*$ amplitudes contributing to both. 
In particular, it was found that the reaction $\gamma^* \to 3\pi$, which builds upon $V \to 3\pi$~\cite{Hoferichter:2014vra, Hoferichter:2018kwz, Hoferichter:2018dmo}, gives the second-largest individual contribution to the HVP integral~\cite{Hoferichter:2019mqg}. The same reaction together with the electromagnetic pion form factor  constrain the doubly virtual pion transition form factor at low virtualities~\cite{Hoferichter:2018kwz, Hoferichter:2018dmo}, which in turn gives the leading contribution to the HLbL process~\cite{Colangelo:2014pva}. Additionally, in the calculation of the helicity partial wave amplitudes for $\gamma^*\gamma^* \to \pi\pi$~\cite{Danilkin:2018qfn, Hoferichter:2019nlq, Danilkin:2019opj}, which are responsible for the two-pion contribution to HLbL, the most important left-hand cut beyond the pion pole is almost exclusively attributed to the $\omega$ exchange. Thus, it depends on the $\omega \to \pi^0 \gamma^*$ TFF. Given the importance of the $\omega \to 3\pi, \pi^0 \gamma^*$ amplitudes and the fact that the currently available ones appear to be at odds with the most recent data, we find it timely to perform a new study of  these reactions. 

The paper is organized as follows. In Section~\ref{sec:formalism} we briefly review the KT formalism for the $\omega \to 3\pi$ decay, and show its relation to the $\omega\pi^0$ TFF. In Section~\ref{sec:results} we discuss fits to the BESIII, MAMI and NA60 data. Our conclusions are given in Section~\ref{sec:conclusions}. Details of the statistical analysis performed to determine uncertainties of the fits are given in the Appendix~\ref{sec:statistics}. 

\section{Formalism}\label{sec:formalism}

\subsection{Kinematics and initial definitions}\label{subsec:kin}
We start by introducing the kinematical definitions for the $\omega(p_{V})$ $\to\pi^0(p_0) \; \pi^+(p_+) \; \pi^-(p_-)$ process. The Mandelstam variables are defined as:
\begin{equation}
    s  = (p_+ + p_-)^2\,,\quad t = (p_0 + p_+)^2\,,\quad u  = (p_0 + p_-)^2\,,
\end{equation}
with $s+t+u=m_{\omega}^{2}+3m_{\pi}^{2}$.
Throughout this manuscript we work in the isospin limit with $m_{\pi}^{2}=m_{\pi^{\pm}}^{2}=m_{\pi^{0}}^{2}$. 
The scattering angle in the $s$-channel, defined by the center of mass of the $\pi^+\pi^-$ pair, is denoted by $\theta_s$ and it is given by:
\begin{equation}\label{Eq:CosTheta}
    \cos\theta_s(s,t,u) = \frac{t - u}{4 \, p(s) \,q(s)}\,,\quad \sin\theta_s(s,t,u)=\frac{\sqrt{\phi(s,t,u)}}{2\sqrt{s}\,p(s) \, q(s)}\,,
\end{equation}
where the momenta $p(s)$ and $q(s)$,
\begin{equation}
   p(s)= \frac{\lambda^{\frac{1}{2}}(s,m_{\pi}^2,m_{\pi}^2)}{2\sqrt{s}}\,,\quad q(s) = \frac{\lambda^{\frac{1}{2}}(s,m_\omega^2,m_{\pi}^2)}{2\sqrt{s}}\,,
\end{equation}
are those of the $\pi^\pm$ and $\pi^0$, respectively, in the $s$-channel. The well-known K\"allen or triangle function $\lambda(a,b,c)$ is defined as~\cite{Kallen:1964lxa}:
\begin{equation}
    \lambda(a,b,c) = a^2 + b^2 + c^2 - 2ab-2bc-2ca~.
\end{equation}
The also well-known Kibble function $\phi(s,t,u)$ is given by~\cite{Kibble:1960zz}:
\begin{equation}
    \phi(s,t,u)=(2\sqrt{s} \; \sin\theta_{s} \; p(s) \, q(s))^{2}=s\,t\,u - m_{\pi}^2 (m_\omega^2 - m_{\pi}^2)^2~,
    \label{Eq:KibbleFunction}
\end{equation}
and it defines the boundaries of the physical regions of the process through the solutions of $\phi(s,t,u)=0$.
The Dalitz-plot boundaries in $t$ for a given value of $s$ for the $\omega\to3\pi$ decay process lie within the interval $[t_{-}(s), \; t_{+}(s)]$, with 
\begin{equation}
    t_{\pm}(s)= \frac{m_{\omega}^{2}+3m_{\pi}^{2}-s}{2} \pm  2 \, p(s) \, q(s)~,
\label{Eq:tmaxmin}
\end{equation}
while the allowed range for $s$ is:
\begin{eqnarray}
    s_{\rm{min}}=4m_{\pi}^{2}\quad \text{to} \quad  s_{\rm{max}}=(m_{\omega}-m_{\pi})^{2}\,.
\end{eqnarray}

\subsection[$\omega \to 3\pi$ amplitude from Khuri--Treiman equations]{\boldmath $\omega \to 3\pi$ amplitude from Khuri--Treiman equations}\label{subsec:KT}

We briefly review here the KT formalism for the $\omega \to 3\pi$ decay amplitudes, refering to Refs.~\cite{Niecknig:2012sj,Danilkin:2014cra} for further details. In the case of vector meson decay into three pions, the helicity amplitude $\mathcal{H}_{\lambda}(s,t,u)$ can be expressed in terms of the single invariant amplitude $F(s,t,u)$,
\begin{equation}
\mathcal{H}_{\lambda}(s,t,u)=i\,\epsilon_{\mu\nu\alpha\beta}\;\epsilon^{\mu}(p_{V},\lambda)\,p_{+}^{\nu}\,p_{-}^{\alpha}\,p_{0}^{\beta}\,\,F(s,t,u)\,,
\label{Eq:AmplitudeOmega3Pi}
\end{equation}
and at the same time decomposed into partial wave amplitudes
\begin{align}
\mathcal{H}_{\lambda}(s,t,u)&=\sum_{J\,\rm{odd}}^{\infty}(2J+1)\;d_{\lambda0}^{J}(\theta_{s})\;h^{(J)}_{\lambda}(s)\,,
\label{Eq:AmpPW}
\end{align}
where $\epsilon_{\mu\nu\alpha\beta}$ is the Levi-Civita tensor, $\epsilon^{\mu}(p_{V},\lambda)$ is the polarization vector of the $\omega$ meson with helicity $\lambda$, and $d_{\lambda0}^{J}(\theta_{s})$ are the Wigner $d$-functions with $\theta_{s}$ given by Eq.~\eqref{Eq:CosTheta}. For a $V \to 3\pi$ decay, $\mathcal{H}_0=0$ and $\mathcal{H}_+ = \mathcal{H}_-$, due to parity. As discussed in Ref.~\cite{Danilkin:2014cra}, one can rewrite the partial wave expansion for the invariant amplitude $F(s,t,u)$ in the following form
\begin{equation}
F(s,t,u)=\sum_{J\,\rm{odd}}^{\infty}(p(s) \, q(s))^{J-1} \; P_{J}^{\prime}(\cos\theta_{s}) \; f_{J}(s)\,,
\label{Eq:AmplitudeF}
\end{equation}
where the exact relation between $h^{(J)}_{+}(s)$ and the kinematic-singularity-free isobar amplitudes $f_{J}(s)$ can be found in Ref.~\cite{Danilkin:2014cra}. The KT representation of the invariant amplitude $F(s,t,u)$ in Eq.~\eqref{Eq:AmplitudeF} consists in substituting the infinite sum of partial waves in the $s$-channel by three finite sums of so-called isobar amplitudes, one for each of the $s$-, $t$- and $u$-channels. By truncating each sum at $J_\text{max}=1$ we obtain the crossing-symmetric isobar decomposition~\cite{Niecknig:2012sj,Danilkin:2014cra,Albaladejo:2019huw}:
\begin{equation}
F(s,t,u)=F(s)+F(t)+F(u)\,, 
\label{Eq:KTdecomposition}
\end{equation}
where each isobar amplitude, $F(x)$, has only right-hand or unitary cut in its respective Mandelstam variable. For the $\pi\pi$ scattering a similar decomposition is known as the reconstruction theorem~\cite{Stern:1993rg, Knecht:1995tr, Albaladejo:2018gif}. For $J=1$, the  relation between $F(s)$ and $f_1(s)$ is obtained by projecting Eq.~\eqref{Eq:KTdecomposition} onto the $s$-channel partial wave,
\begin{align}
\label{Eq:f1}
f_1(s)& = F(s)+\hat{F}(s)\,, \\
\hat{F}(s)& \equiv 3\int_{-1}^{1}\frac{dz_{s}}{2} \; (1-z_{s}^{2}) \; F(t(s,z_{s}))\,,
\label{Eq:Fhat}
\end{align}
where the so-called inhomogeneity $\hat{F}(s)$ contains the $s$-channel projection of the left-hand cut contributions due to the $t$- and $u$-channels. Its evaluation in the decay region requires a proper analytical continuation~\cite{Bronzan:1963mby}. Assuming elastic unitarity with only two-pion intermediate states, we arrive at the KT equation for the $\omega\to3\pi$ decay, \ie the
unitarity relation for the isobar amplitude $F(s)$:
\begin{align}
\label{Eq:discOmegaPiSingleVariable}
    {\rm{disc}}\,F(s)&=2i\left(F(s)+\hat{F}(s)\right) \; \sin\delta(s) \; e^{-i\delta(s)} \; \theta(s-4m_{\pi}^{2})\,,
\end{align}
where $\delta(s)$ is the $P$-wave $\pi\pi$ phase shift. 
Given the discontinuity relation Eq.\,(\ref{Eq:discOmegaPiSingleVariable}), one can write an unsubtracted dispersion relation (DR) for $F(s)$ as
\begin{align}
\label{Eq:DRforF(s)}
    F(s)=\frac{1}{2\pi i}\int_{4m_{\pi}^2}^{\infty}d s' \; \frac{{\rm{disc}}\,F(s')}{s'-s}\,,
\end{align}
which can be solved numerically~\cite{Niecknig:2012sj,Guo:2014vya,Albaladejo:2017hhj,Gasser:2018qtg}. Its solution is given in terms of the usual Omn\`{e}s function~\cite{Omnes:1958hv},
\begin{equation}
    \Omega(s)=\exp\left[\frac{s}{\pi}\int_{4m_{\pi}^{2}}^{\infty}\frac{ds^{\prime}}{s^{\prime}}\frac{\delta(s^{\prime})}{s^{\prime}-s}\right]\,,
\label{Eq:Omnes}
\end{equation}
defined by the real phase shift $\delta(s)$. For the latter we take the solution of the Roy equations of Ref.\,\cite{GarciaMartin:2011cn}, that are valid roughly up to 1.3 GeV. From 1.3 GeV on we smoothly guide $\delta(s)$ to $\pi$ to obtain the expected asymptotic $1/s$ fall-off behavior for the pion vector form factor (see \eg Ref.~\cite{Gonzalez-Solis:2019iod}). The solution of Eq.\,(\ref{Eq:DRforF(s)}) is written as:
\begin{equation}
F(s)=\Omega(s)\left(a+\frac{s}{\pi}\int_{4m_{\pi}^{2}}^{\infty}\frac{ds^{\prime}}{s'}\frac{\sin\delta(s^{\prime}) \, \hat{F}(s^{\prime})}{|\Omega(s^{\prime})|\left(s^{\prime}-s\right)}\right)\,,
\label{Eq:KTGeneral}
\end{equation}
where the (complex) normalization constant $a=|a|\,e^{i\phi_{a}}$ is an overall normalization of the amplitude and can be factored out. Using PDG data, $|a|$ can be fixed to reproduce the experimental $\omega\to3\pi$ decay width. No observables of the decay are sensitive to the overall phase $\phi_a$. Due to the asymptotic behavior of $F(s)$ implied by Eq.~\eqref{Eq:KTGeneral}, the amplitude $F(s,t,u)$ satisfies the Froissart-Martin bound~\cite{Froissart:1961ux,Martin:1962rt,Niecknig:2012sj}.

We emphasize that, even though $F(s)/\Omega(s)$ in Eq.~\eqref{Eq:KTGeneral} looks like a once-subtracted dispersion relation, $F(s)$ actually satisfies the unsubtracted dispersion relation given in Eq.~\eqref{Eq:DRforF(s)}. Therefore, the energy dependence of $F(s)$ is a pure prediction, which in the elastic approximation is given solely by the $P$-wave $\pi\pi$ phase shift. Note that Eq.~\eqref{Eq:KTGeneral} can be written in the form 
\begin{equation}
F(s)=\Omega(s)\left(a+b'\,s+\frac{s^2}{\pi}\int_{4m_{\pi}^{2}}^{\infty}\frac{ds^{\prime}}{(s')^2}\frac{\sin\delta(s^{\prime}) \, \hat{F}(s^{\prime})}{|\Omega(s^{\prime})|\left(s^{\prime}-s\right)}\right)\,,
\label{Eq:KTGeneral_2}
\end{equation}
if $b'$ satisfies the following sum rule~\cite{Niecknig:2012sj}:
\begin{equation}\label{Eq:SumRuleAn}
b\equiv b'/a=\frac{1}{\pi}\int_{4m_{\pi}^{2}}^{\infty}\frac{ds^{\prime}}{(s')^2}\frac{\sin\delta(s^{\prime}) \,\hat{F}(s^{\prime})/a}{|\Omega(s^{\prime})|}\,.
\end{equation}
In Ref.~\cite{Niecknig:2012sj} its value was computed, with the result:
\begin{equation}\label{Eq:SumRuleNum}
b_\text{sum} \simeq\, 0.55\,e^{0.15\,i}\ \text{GeV}^{-2}~,
\end{equation}
which we reproduce as a numerical cross-check. We note that, due to the three-particle cut, which become physically accessible in the decay amplitude, this subtraction constant is complex and is thus determined by two parameters, its modulus and phase, $b=|b|\,e^{i\phi_{b}}$. 

\begin{figure}\centering
\includegraphics{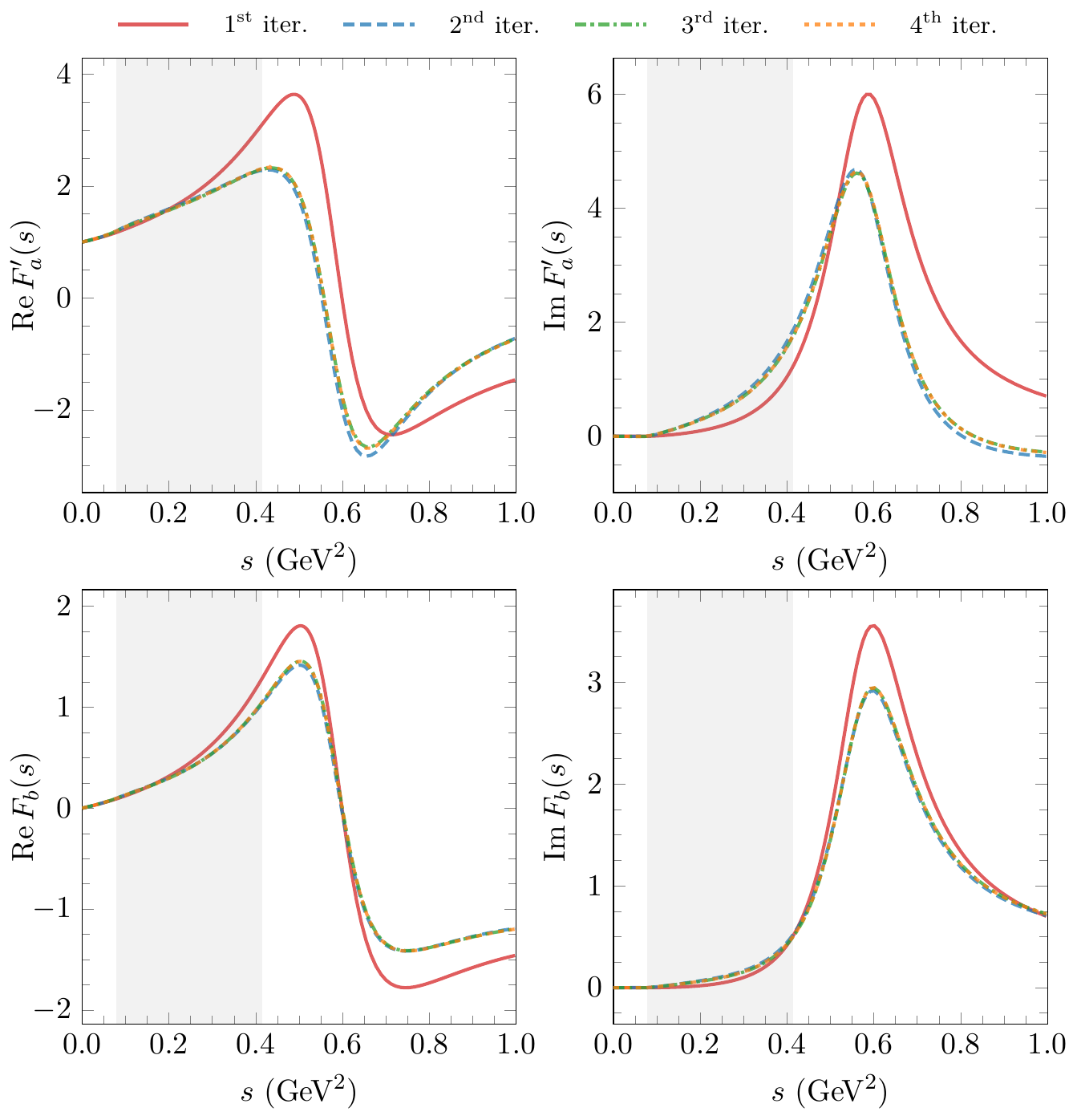}
\caption{\label{Fig:IterationsFaFb}Convergence behavior of the iterative procedure for the real (left plots) and imaginary (right plots) parts of the amplitudes $F'_{a}(s)$ (Eq.~\eqref{Eq:Fa}, upper plots) and $F_{b}(s)$ (Eq.~\eqref{Eq:Fb}, lower plots). The shaded area corresponds to the $\omega \to 3\pi$ physical decay region.}
\end{figure}

In contrast to the unsubtracted DR in Eq.~\eqref{Eq:DRforF(s)}, one can start from a once-subtracted DR:
\begin{equation}
\label{Eq:DRforF(s)_once}
    F(s)=F(0) + \frac{s}{2\pi i}\int_{4m_{\pi}^2}^{\infty}d s' \; \frac{{\rm{disc}}\,F(s')}{s'(s'-s)}~.
\end{equation}
The solution to Eq.~\eqref{Eq:DRforF(s)_once} can be constructed as the linear combination~\cite{Niecknig:2012sj,Albaladejo:2017hhj}:
\begin{subequations}\label{eqs:ourFs}
\begin{equation}
    F(s)=a\left[F_{a}'(s)+b\,F_{b}(s)\right]\,,
\label{Eq:KTtwosub}
\end{equation}
where now $b$ is not constrained to satisfy Eq.~\eqref{Eq:SumRuleAn}, and the functions $F_{a}'(s)$ and $F_{b}(s)$ are given by
\begin{eqnarray}
    F_{a}'(s)&=&\Omega(s)\left[1+\frac{s^{2}}{\pi}\int_{4m_{\pi}^{2}}^{\infty}\frac{ds^{\prime}}{s^{\prime2}}\frac{\sin\delta(s^{\prime}) \, \hat{F}'_{a}(s^{\prime})}{|\Omega(s^{\prime})|(s^{\prime}-s)}\right]\,,\label{Eq:Fa}\\[1ex]
    F_{b}(s)&=&\Omega(s)\left[s+\frac{s^{2}}{\pi}\int_{4m_{\pi}^{2}}^{\infty}\frac{ds^{\prime}}{s^{\prime2}}\frac{\sin\delta(s^{\prime}) \, \hat{F}_{b}(s^{\prime})}{|\Omega(s^{\prime})|(s^{\prime}-s)}\right]\,.
\label{Eq:Fb}
\end{eqnarray}
\end{subequations}
These functions only need to be calculated once, since they are independent of the numerical values of $a$ and $b$, which become fit parameters, as will be discussed in Sec.~\ref{sec:results}. For completeness, in Fig.~\ref{Fig:IterationsFaFb} we show the solutions for $F^\prime_{a}(s)$ and $ F_{b}(s)$ using a numerical iterative procedure similar to those employed in previous works~\cite{Kambor:1995yc,Anisovich:1996tx,Danilkin:2014cra,Albaladejo:2017hhj}. 

By introducing one subtraction we reduce the sensitivity to the unknown high energy behavior of the phase shift and/or to the inelastic contributions, which are thus embeded in the subtraction constant. Furthermore, the parameter $b$ allows to parametrize some unknown energy dependence of the $\omega \to 3\pi$ interaction not directly related to $\pi\pi$ rescattering.\footnote{For instance, in Refs.~\cite{Anisovich:1996tx,Descotes-Genon:2014tla,Albaladejo:2017hhj}, in the context of $\eta \to 3\pi$ KT equations, the subtraction constants are used to match the dispersive amplitude and its derivatives to the chiral ones, thus constraining the value of those parameters.} Strictly speaking, the amplitude $F(s,t,u)$ built from $F(s)$ in Eq.~\eqref{Eq:KTtwosub} would not satisfy the Froissart-Martin bound~\cite{Froissart:1961ux,Martin:1962rt,Niecknig:2012sj} for an arbitrary value of the parameter $b \neq b_\text{sum}$ [\cf Eq.~\eqref{Eq:SumRuleAn}]. In practice, however, given the low-energy regime in which Eq.~\eqref{Eq:KTtwosub} is applied, this bound is not relevant and we therefore do not constrain the value of $b$.

Finally, the measured differential decay width can be written in terms of the invariant amplitude $F(s,t,u)$ as
\begin{equation}
\frac{d^2 \Gamma}{d s\,d t} = \frac{1}{(2\pi)^3}\, \frac{1}{32\,m_\omega^3} \, \frac{1}{3} \frac{\phi(s,t,u)}{4}  \; | F(s,t,u) |^2\,.
\label{Eq:DecayWidth}
\end{equation}
The $\omega \to 3\pi$ Dalitz plot distribution is conventionally  parametrized in terms of  the variables $X$, $Y$ defined by 
\begin{equation}\label{eq:dalitzxy}
X=\frac{t-u}{\sqrt{3}R_{\omega}}\,,\quad Y=\frac{s_{c}-s}{R_{\omega}}\,,
\end{equation}
where $s_{c}=\frac{1}{3}(m_{\omega}^{2}+3m_{\pi}^{2})$ and $R_{\omega}=\frac{2}{3}m_{\omega}(m_{\omega}-3m_{\pi})$. The $\{X,Y\}$ variables are related 
 to  the polar ones $\{Z, \varphi\}$ through $X=\sqrt{Z}\,\cos\varphi$ and $Y=\sqrt{Z}\,\sin\varphi$, which enter into the Dalitz-plot expansion as:
\begin{equation}
|F_{\rm{pol}}(Z,\varphi)|^{2}=|N|^{2}\left[1+2\alpha Z+2\beta Z^{3/2}\sin3\varphi+2\gamma Z^{2}+\mathcal{O}(Z^{5/2})\right]\,.
\label{PolynomialRepresentation}
\end{equation}
In Eq.~\eqref{PolynomialRepresentation}, $\alpha,\beta$ and $\gamma$ are the real-valued Dalitz-plot parameters and $N$ is an overall normalization. In order to obtain $\alpha$, $\beta$, and $\gamma$ for a given theoretical amplitude $F_\text{th}(z,\phi)$ we minimize~\cite{Niecknig:2012sj}
\begin{eqnarray}\label{Eq:chi2Omega3Pi}
\xi_{\rm{Dalitz}}^{2}&=&\frac{1}{N_{D}}\int_{D}dZ\,d\varphi\left[\frac{\phi(Z,\varphi)}{\phi(0,0)}\frac{|F_{\rm{pol}}(Z,\varphi)|^{2}-|F_{\rm{th}}(Z,\varphi)|^{2}}{|N|^{2}}\right]^{2}\,,\\[1ex]
N_{D}&=&\int_{D}dZ\,d\varphi\,,\nonumber
\end{eqnarray}
where $D$ is the area of the Dalitz plot, $\phi(Z,\varphi)$ is $\phi(s,t,u)$ with $s$, $t$, and $u$ expressed in terms of the polar variables $\{Z,\varphi\}$, and $\xi_{\rm{Dalitz}}^{2}$ denotes the average deviation of the theoretical description and the polynomial one relative to the Dalitz plot center. We also note that the Dalitz-plot parameters enter into the difference in Eq.~\eqref{Eq:chi2Omega3Pi} linearly, and thus the minimization can be algebraically solved.

\subsection[$\omega\pi^{0}$ transition form factor]{\boldmath $\omega\pi^{0}$ transition form factor}\label{subsec:TFF}

\begin{figure}\centering
\includegraphics[scale=1.3]{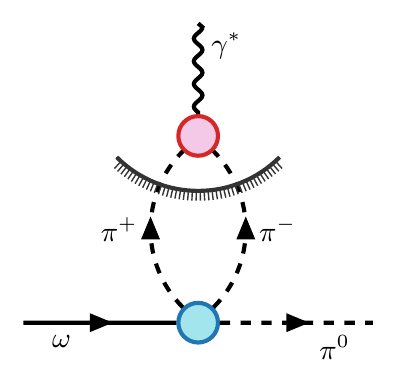}
\caption{Diagrammatic representation of the two-pion contribution to the discontinuity of the $\omega\pi^0$ transition form factor [\cf Eq.\,(\ref{Eq:Discontinuity})].
The blue and red circles represent, respectively, the full $s$-channel $P$-wave $\omega\to3\pi$ amplitude $f_{1}(s)$ and the pion vector form factor $F_{\pi}^{V}(s)$.\label{fig:tff}}
\end{figure}

The $\omega\pi^0$ transition form factor, $f_{\omega\pi^0}(s)$, controls the $\omega \to \pi^0 \gamma^{\ast}$ amplitude, see \eg Refs.~\cite{Schneider:2012ez,Danilkin:2014cra}. A dispersive representation of $f_{\omega\pi^{0}}(s)$ is fully determined, up to possible subtractions, by the discontinuity across the right hand cut. In order to be consistent with the elastic approximation in the $\omega \to 3\pi$ study, we include only the two-pion contribution to the discontinuity (see Fig.~\ref{fig:tff} for a diagrammatic  interpretation)~\cite{Koepp:1974da,Schneider:2012ez}  :
\begin{equation}
{\rm{disc}}f_{\omega\pi^{0}}(s)=i \; \frac{p^{3}(s)}{6\pi\sqrt{s}} \; {F_{\pi}^{V}}^{*}(s) \; f_{1}(s) \; \theta(s-4m_{\pi}^{2})\,,
\label{Eq:Discontinuity}
\end{equation}
which requires as input the full $s$-channel $P$-wave $\omega\to3\pi$ amplitude $f_{1}(s)$ given in Eq.~\eqref{Eq:f1} and the the pion vector form factor $F_{\pi}^{V}(s)$, which we approximate by the Omn\`{e}s function $\Omega(s)$ given in Eq.~\eqref{Eq:Omnes}. This is a reasonable approximation given the low $\omega\pi^0$ invariant mass that we explore in this work. In order to reduce the sensitivity of the dispersive integral to the higher-energy region, we use a once-subtracted dispersion relation \begin{eqnarray}
f_{\omega\pi^{0}}(s)=|f_{\omega\pi^{0}}(0)|\,e^{i\phi_{\omega\pi^{0}}(0)}+\frac{s}{12\pi^{2}}\int_{4m_{\pi}^{2}}^{\infty}\frac{ds^{\prime}}{(s^{\prime})^{3/2}}\frac{p^{3}(s^{\prime}) \; {F_{\pi}^{V}}^*(s^{\prime}) \; f_{1}(s^{\prime})}{(s^{\prime}-s)}\,,
\label{Eq:OmegaPiFF1sub}
\end{eqnarray}
where we indicate explicitly the existence of a non-vanishing phase of $f_{\omega\pi^{0}}(s)$ at $s=0$. This is implied by the cross-channel effects, \ie the functions $F_{\pi}^{V}(s)$ and $f_{1}(s)$ do not have the same phase, and the discontinuity of $f_{\omega\pi^{0}}(s)$ is in general complex~\cite{Schneider:2012ez}, even for $\phi_a=0$. The modulus of the subtraction constant  $|f_{\omega\pi^{0}}(0)|$ can be fixed from the $\omega\to\pi^{0}\gamma$ partial decay width 
\begin{equation}
\Gamma(\omega\to\pi^{0}\gamma)=\frac{e^{2}(m_{\omega}^{2}-m_{\pi^{0}}^{2})^{3}}{96\pi m_{\omega}^{3}}\; |f_{\omega\pi^{0}}(0)|^{2}\,,
\label{Eq:OmegaPiGammaWidth}
\end{equation}
while its phase $\phi_{\omega\pi^{0}}(0)$ is a free parameter that will be fixed from fits to the transition form factor experimental data. On the other hand, this phase appears only in the first term of Eq.~\eqref{Eq:OmegaPiFF1sub}, while the phase $\phi_a$ appears only in the second term. Thus, only the relative phase $\phaf - \phi_a$ is relevant, and, bearing this in mind, we set $\phi_a = 0$.

\section{Results}\label{sec:results}
\begin{table}\centering\begin{tabular}{|c|c|c|c|c|} \hline
                        &  Reference & $\alpha \times 10^3$     & $\beta \times 10^3$     & $\gamma \times 10^3$ \\ \hline \hline
\multirow{7}{*}{\begin{tabular}{c} 2 par.\\ $(\alpha,\beta)$ \end{tabular}}
& Ref.~\cite{Terschlusen:2013iqa} ($\pi\pi$ rescattering)      
                                          & $190$ & $54$ & -- \\ \cline{2-5}
& Ref.~\cite{Danilkin:2014cra}, w KT      & $ 84$ & $28$ & -- \\
& Ref.~\cite{Danilkin:2014cra}, w/o KT    & $125$ & $30$ & -- \\  \cline{2-5}
& Ref.~\cite{Niecknig:2012sj}, w KT       & $ 79(5)$ & $26(2)$ & -- \\
& Ref.~\cite{Niecknig:2012sj}, w/o KT     & $130(5)$ & $31(2)$ & -- \\  \cline{2-5}
& WASA-at-COSY~\cite{Adlarson:2016wkw}    & $133(41)$ & $37(54)$ & -- \\
& BESIII~\cite{Ablikim:2018yen}           & $120.2(8.1)$ & $29.5(9.6)$ & -- \\ \cline{2-5}          
& This work, low $\phi_{\omega\pi^0}(0)$  & $121.2(7.7)$ & $25.7(3.3)$ & --  \\
& This work, high $\phi_{\omega\pi^0}(0)$ & $120.1(7.7)$ & $30.2(4.3)$ & --  \\  \hline \hline
\multirow{7}{*}{\begin{tabular}{c} 3 par.\\ $(\alpha,\beta,\gamma)$ \end{tabular}}
& Ref.~\cite{Terschlusen:2013iqa} ($\pi\pi$ rescattering)      
                                          & $172$ & $43$ & $50$ \\ \cline{2-5}
& Ref.~\cite{Danilkin:2014cra}, w KT      & $ 80$ & $27$ & $8$ \\
& Ref.~\cite{Danilkin:2014cra}, w/o KT    & $113$ & $27$ & $24$ \\   \cline{2-5}
& Ref.~\cite{Niecknig:2012sj}, w KT       &  $77(4)$  & $26(2)$  &  $5(2)$ \\
& Ref.~\cite{Niecknig:2012sj}, w/o KT     & $116(4)$  & $28(2)$  & $16(2)$ \\   \cline{2-5}
& BESIII~\cite{Ablikim:2018yen}           & $111(18)$ & $25(10)$ & $22(29)$ \\ \cline{2-5}
& This work, low $\phi_{\omega\pi^0}(0)$  & $112(15)$ & $23(6) $ & $29(6)$ \\  
& This work, high $\phi_{\omega\pi^0}(0)$ & $109(14)$ & $26(6) $ & $19(5)$ \\ \hline
\end{tabular}\caption{Dalitz plot parameters $\alpha$, $\beta$, and $\gamma$, obtained by previous theoretical~\cite{Terschlusen:2013iqa,Niecknig:2012sj,Danilkin:2014cra} and experimental~\cite{Adlarson:2016wkw,Ablikim:2018yen} analyses. For the dispersive analyses~\cite{Niecknig:2012sj,Danilkin:2014cra}, we show the results obtained with and without KT equations (\ie, with $F(s)$ proportional to an Omn\'es function, see also Subsec.~\ref{subsec:otherapproaches}). Also shown are our results, for the two solutions that we find in this work. The upper (lower) part of the Table show the results when 2 (3) Dalitz plot parameters are determined.
\label{table:DalitzPlot}}\end{table}

\subsection{General approach}\label{subsec:general}
The two amplitudes defined in the previous section depend on a total of 
five real parameters. The $\omega \to 3\pi$ amplitude depends on $|a|$ and $b = |b| \exp(i \phi_b)$ [\cf Eq.~(\ref{Eq:KTtwosub})], whereas the $\omega \pi^0$ transition form factor additionally depends on the subtraction constant at $s=0$, $f_{\omega \pi^0}(0)$, also complex. To fix those unknown constants we will use the following experimental information:
\begin{enumerate}[a)]
\item the recent determination of the $\omega \to 3\pi$ decay Dalitz plot parameters by BESIII~\cite{Ablikim:2018yen}, shown in Table~\ref{table:DalitzPlot}. We note that there are two different determinations, labeled as ``2 par.'' and ``3 par.'', corresponding to whether the Dalitz plot distribution is assumed to be described by two ($\alpha$ and $\beta$) or three ($\alpha$, $\beta$, and $\gamma$) parameters,  respectively;
\item the $\omega \to 3\pi$ and $\omega \to \pi^0 \gamma$ decay widths, for which we take the PDG values~\cite{Tanabashi:2018oca}, $\Gamma_\omega = 8.49 \pm 0.08\ \text{MeV}$, $\mathcal{B}(\omega \to 3\pi) = 89.3 \pm 0.6\ \%$, and $\mathcal{B}(\omega \to \pi^0 \gamma) = 8.40 \pm 0.22\ \%$;
\item the data on $\left \lvert f_{\omega \pi^0}(s) / f_{\omega \pi^0}(0) \right \rvert^2$ for low $\omega\pi^0$ invariant mass by the A2 collaboration at MAMI~\cite{Adlarson:2016hpp} and by the NA60 collaboration at SPS~\cite{Arnaldi:2009aa,Arnaldi:2016pzu}. From the NA60 collaboration data, we will only consider for our fits the most up to date analysis~\cite{Arnaldi:2016pzu}.
\end{enumerate}

For each of these sets we define the following $\chi^2$ functions,
\begin{subequations}\label{eq:chi2defs}\begin{align}
\chi^2_\text{DP} & = 
\left( \frac{\alpha^\text{(th)}-\alpha^\text{(exp)}}{\sigma_\alpha} \right)^2 +
\left( \frac{ \beta^\text{(th)}- \beta^\text{(exp)}}{\sigma_\beta}  \right)^2 +
\left( \frac{\gamma^\text{(th)}-\gamma^\text{(exp)}}{\sigma_\gamma} \right)^2~, \label{eq:chi2DP}\\
\chi^2_\Gamma & =
\left( \frac{\Gamma_{\omega \to 3\pi}^\text{(th)}-\Gamma_{\omega \to 3\pi}^\text{(exp)}}{\sigma_{\Gamma_{\omega \to 3\pi}}} \right)^2 +
\left( \frac{\Gamma_{\omega \to \pi^0 \gamma}^\text{(th)}-\Gamma_{\omega \to \pi^0 \gamma}^\text{(exp)}}{\sigma_{\Gamma_{\omega \to \pi^0 \gamma}}} \right)^2~,\\
\chi^2_{\text{A2},\text{NA60}} & = \sum_i \left( \frac{\left\lvert f^\text{(th)}_{\omega\pi}(s_i) \right\rvert^2 - \left\lvert f^{\text{(exp)},\,i}_{\omega\pi} \right\rvert^2}{\sigma_{f_{\omega\pi}^{(i)}}} \right)^2~,
\end{align}\end{subequations}
where in $\chi^2_\text{A2}$ and $\chi^2_\text{NA60}$ the sum runs over the experimental points with $\sqrt{s}_i \leqslant 0.65\ \text{GeV}$. 

\begin{figure}\centering
\includegraphics{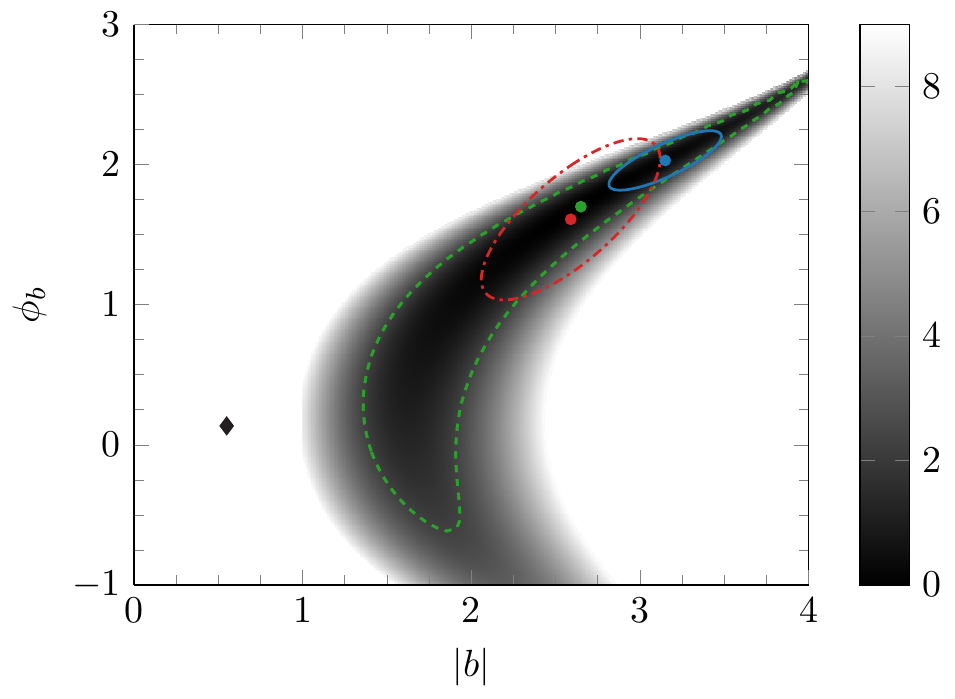}
\caption{Comparison of different determinations of the free parameters $(\modb, \phab)$ for the ``2 par.'' case (similar results are obtained for ``3 par.''). The blue solid (red dashed) line represents our two-parameter $(\modb, \phab)$ $1\sigma$ (68\% CL) error ellipse for the low (high) $\phaf$ solution (global fits, $\overline{\chi}^2$, \cf Eq.~\eqref{eq:barchi2}) described in Subsec.~\ref{subsec:globalfits}, with parameters given in Table~\ref{table:parameters}. Errors are estimated with MC resampling, as explained in the text and in Appendix~\ref{sec:statistics}. The background color represents the value of the function $\chi^2_\text{DP}$ [\cf Eq.~\eqref{eq:chi2DP}] as a function of $\modb$ and $\phab$, thus corresponding to the fit described in Subsec.~\ref{subsec:general} [\cf Eq.~\eqref{eq:fitbes2dp}]. The green dashed line represents the $\chi^2_\text{DP} \simeq 2.3$ contour (at the minimum, $\chi^2_\text{DP}=0$), that corresponds to the two-parameter $1\sigma$ region. Lastly, the black diamond represents the value $b_\text{sum}$, Eq.~\eqref{Eq:SumRuleNum}.\label{fig:sumrule}}
\end{figure}

\begin{figure}\centering
\includegraphics{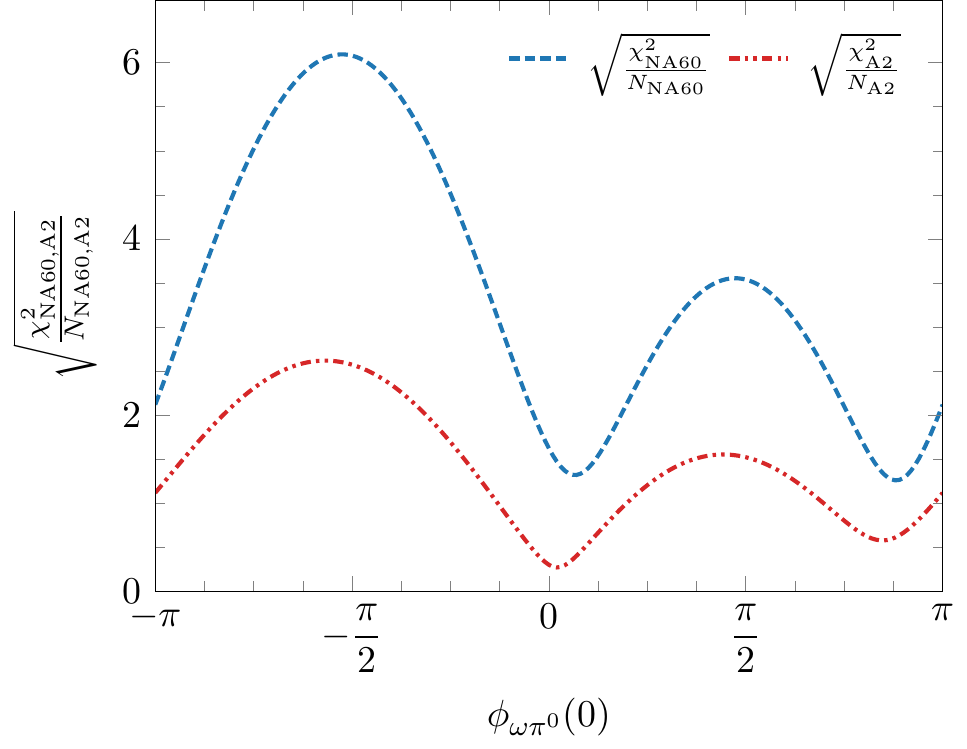}
\caption{Dependence of the $\chi^2_\text{NA60,A2}$ functions on the phase $\phaf$ for fixed values of the other free parameters, as described in the text.\label{fig:chi2run}}
\end{figure}

To determine the role of each data set, we start by considering the $\omega \to 3\pi$ Dalitz plot parameters alone, since they only depend on $\modb$ and $\phi_b$. In a first step, we fix $\modb$ and $\phab$ from the Dalitz plot parameters (\ie, by minimizing $\chi^2_\text{DP}$), and, in a second step, we fix $\moda$ and $\modf$ from the decay widths (\ie, by minimizing $\chi^2_\Gamma$). We obtain the following values for the ``2 par.'' case:
\begin{subequations}\label{eq:fitbes}
\begin{align}
    \modb = 2.65(1.10)\ \text{GeV}^{-2}~,          & \qquad \phab = 1.70^{+1.40}_{-0.70}~, \label{eq:fitbes2dp} \\
    10^{-2} \moda = 2.82(72)\ \text{GeV}^{-3}~,  & \qquad \modf = 2.314(31)\ \text{GeV}^{-1}~, \qquad \text{(``2 par.'')}~, \nonumber
\end{align}
whereas, for the ``3 par.'' case, one gets:
\begin{align}
    \modb = 2.88^{+1.65}_{-0.85}\ \text{GeV}^{-2}~,          & \qquad \phab = 1.85^{+1.45}_{-0.45}~, \label{eq:fitbes3dp}\\
     10^{-2} \moda = 3.00(68)\ \text{GeV}^{-3}~, & \qquad \modf = 2.314(31)\ \text{GeV}^{-1}~, \qquad \text{(``3 par.'')}~. \nonumber
\end{align}
\end{subequations}
Because we are fitting two or three experimental points with two free parameters, the $\chi^2_\text{DP}$ is zero for the ``2 par.'' case, and almost zero for the ``3 par.'' case. In turn, this manifests in the large value of the errors shown in Eqs.~\eqref{eq:fitbes}. These errors are obtained through the condition $\Delta\chi^2_\text{DP} \leqslant 1$. We  note that the value obtained for $b$ is quite different from the value of $b_\text{sum}$ [\cf Eq.~\eqref{Eq:SumRuleNum}], as also shown in Fig.~\ref{fig:sumrule}. This reinforces the idea that, in order to achieve a proper description of the BESIII Dalitz plot parameters, an additional subtraction is needed within the KT formalism.\footnote{We note here that in the $\phi \to 3\pi$ study of Ref.~\cite{Niecknig:2012sj} it is also found that the fitted value of $b$ differs from the equivalent sum rule, although the differences are much smaller than in our $\omega \to 3\pi$ case.}

In Eqs.~\eqref{eq:fitbes} we have fixed all the free parameters but $\phaf$, and we now study the dependence of the $\omega\pi^0$ TFF on this phase. In Fig.~\ref{fig:chi2run} we show how $\chi^2_\text{A2}$ and $\chi^2_\text{NA60}$ depend on this phase for fixed values of the other   parameters. We present the result for the ``3 par.'' case, Eq.~\eqref{eq:fitbes3dp}, but an analogous result is obtained for the ``2 par.'' case. We observe that there are two minima, one at $\phaf \simeq 0.2$ and another one at $\phaf \simeq 2.5$, to which we refer in what follows as ``low $\phaf$'' and ``high $\phaf$'' solutions, respectively. Furthermore, it is observed that the values of $\chi^2_\text{NA60,A2}$ are similar in both cases, \ie, both solutions describe the data with similar quality.

\subsection{Global fit results}\label{subsec:globalfits}
\begin{table}\centering
\begin{tabular}{|r|c|c|c|c|} \hline
 & \multicolumn{2}{c|}{2 par.} & \multicolumn{2}{c|}{3 par.} \\ \hline
 & low $\phaf$  & high $\phaf$ & low $\phaf$ & high $\phaf$ \\ \hline\hline
$10^{-2} \moda$ $\left[ \text{GeV}^{-3} \right]$ 
                                 & $3.14(25)$   & $2.63(25)$  & $3.11(28)$  & $2.70(30)$ \\
$\modb\ \left[\text{GeV}^{-2} \right]$        & $3.15(22)$   & $2.59(35)$  & $3.25(26)$  & $2.65(35)$  \\
$\phab$                          & $2.03(14)$   & $1.61(38)$  & $2.03(13)$  & $1.70(27)$  \\
$\modf$ $\left[ \text{GeV}^{-1} \right]$         
                                 & $2.314(32)$  & $2.314(32)$ & $2.314(32)$ & $2.315(32)$ \\
$\phaf$                          & $0.207(60)$  & $2.39(46)$  & $0.195(76)$ & $2.48(31)$  \\ \hline
$\chi^2_\text{DP}$ $[ N_\text{DP}=2\text{ or }3]$        & $0.19$ & $<0.01$ & $0.10$ & $0.03$\\
$10^4 \chi^2_\Gamma$ $[ N_\Gamma=2]$        & $2.4$  & $2.4$ & $1.1$ & $3.5$ \\
$\chi^2_\text{A2}$ $[ N_\text{A2}=14]$      & $2.3$  & $3.6$ & $2.4$ & $3.7$ \\
$\chi^2_\text{NA60}$  $[ N_\text{NA60}=22]$ & $31$   & $35$  & $31$ & $35$ \\ \hline
\end{tabular}
\caption{Values of the fitted parameters (upper part) and of the different $\chi^2$ functions (lower part) for the four different fits considered in this work (see Subsec.~\ref{subsec:globalfits} for details). The errors represent our $1\sigma$ uncertainties, and are computed through MC resampling, as explained in the text and in the Appendix~\ref{sec:statistics}.
\label{table:parameters}}
\end{table}

\begin{figure}\centering
\includegraphics{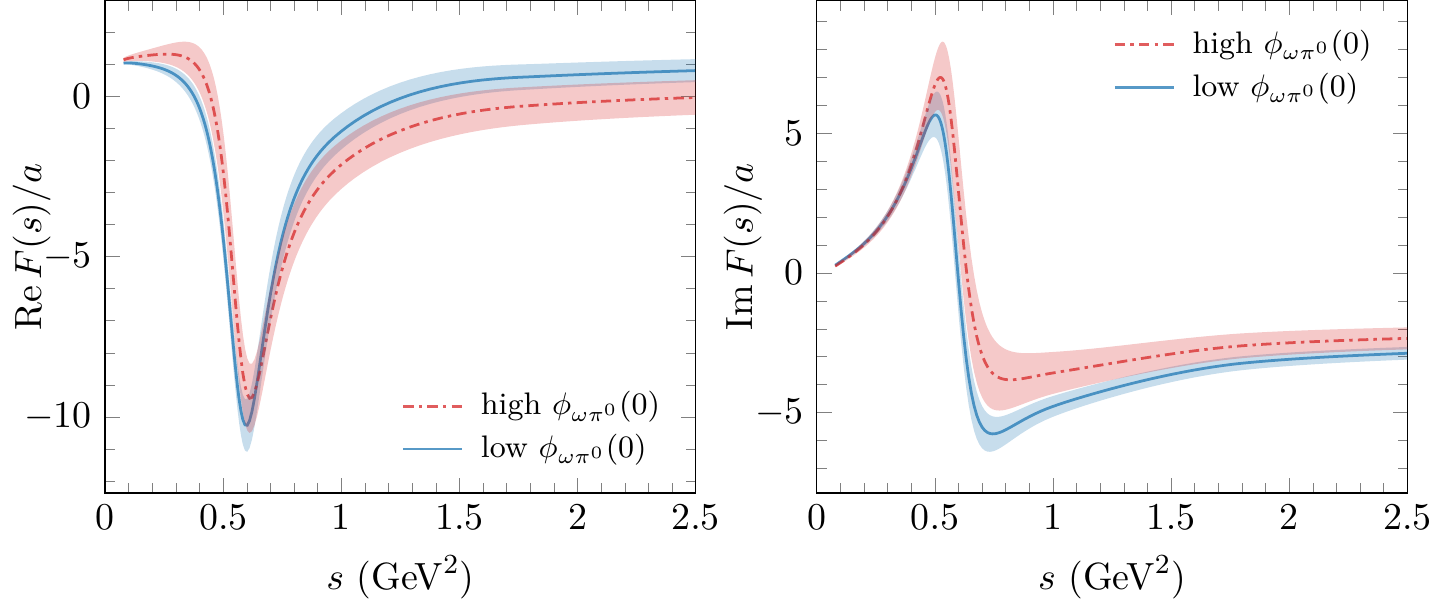}
\caption{Real (left) and imaginary (right) parts of the function $F(s)$ (modulo $\moda$) [\cf Eq.~\eqref{eqs:ourFs}], with $b=\modb e^{\phab}$ as in Table~\ref{table:parameters}, case ``3 par.''. We show the results for the ``low $\phaf$'' (blue solid) and ``high $\phaf$'' (red dash-dotted) solutions. The error bands are obtained from a MC analysis of the fitted data, and represent the correlated $1\sigma$ uncertainty in our parameters.\label{fig:fig_reim_fs}}
\end{figure}

Given that we are able to separately reproduce the  experimental data on the two reactions, in the next step  we perform a simultaneous fit. To that end, we minimize the following $\chi^2$-like function,
\begin{equation}\label{eq:barchi2}
\overline{\chi}^2 = N \left( \frac{\chi^2_\text{DP}}{N_\text{DP}} + \frac{\chi^2_\Gamma}{N_\Gamma} + \frac{\chi^2_\text{NA60}}{N_\text{NA60}} + \frac{\chi^2_\text{A2}}{N_\text{A2}} \right)~,
\end{equation}
where $N_\text{DP} = 2\text{ or }3$ is the number of Dalitz plot parameters considered, $N_\Gamma = 2$ the experimental partial widths, $N_\text{A2} =14$ and $N_\text{NA60} = 22$ the experimental points in the two sets for $\fwsq$, and $N=N_\text{DP} + N_\Gamma + N_\text{A2} + N_\text{NA60}$. This ensures that $\chi^2$ functions with a smaller number of points are well represented in $\overline{\chi}^2$, and are not overriden by those with a larger number of points.

   When the simultaneous fit is performed we observe, as expected, that the two solutions remain. The two minima are well separated, as can be seen in Fig.~\ref{fig:chi2run}, so that we can analyze each solution individually. Besides these two solutions, we must also consider the two different sets of Dalitz-plot parameters given by the BESIII collaboration, as shown in Table~\ref{table:DalitzPlot}. Therefore, we perform four different fits, and the fitted parameters, as well as the individual values of the $\chi^2$ functions, are compiled in Table~\ref{table:parameters}. The quoted errors are obtained through a Monte Carlo (MC) analysis with data resampling (bootstrap~\cite{recipes,EfroTibs93,Landay:2016cjw}), and they represent  $1\sigma$ level uncertainties (see Appendix~\ref{sec:statistics} for further details). The values obtained for the individual $\chi^2$ functions imply a good quality of the fits. As a consistency check between the ``2 par.'' and ``3 par.'' data sets, we note that the values of the parameters are similar among the two ``low $\phaf$'' solutions (second and fourth columns in Table~\ref{table:parameters}), as well as among the two ``high $\phaf$'' solutions (third and fifth columns). As an illustration, we show in Fig.~\ref{fig:fig_reim_fs} the function $F(s)$ obtained using the values of the parameters that correspond to the ``3 par.'' set, for both solutions. Regarding specifically the values of $\modb$ and $\phab$, we note that both solutions fall well within the region determined by the fit to only BESIII data described in Subsec.~\ref{subsec:general}, see Fig.~\ref{fig:sumrule}. This means that both solutions originate from that, but have much more constrained uncertainties as a result of the inclusion of the TFF data. We also note that the two widths considered in the $\chi^2$ ($\Gamma_{\omega \to 3\pi}$ and $\Gamma_{\omega \to \pi^0 \gamma}$) are reproduced with the same central values and errors as the experimental ones.

\begin{figure}\centering
\includegraphics{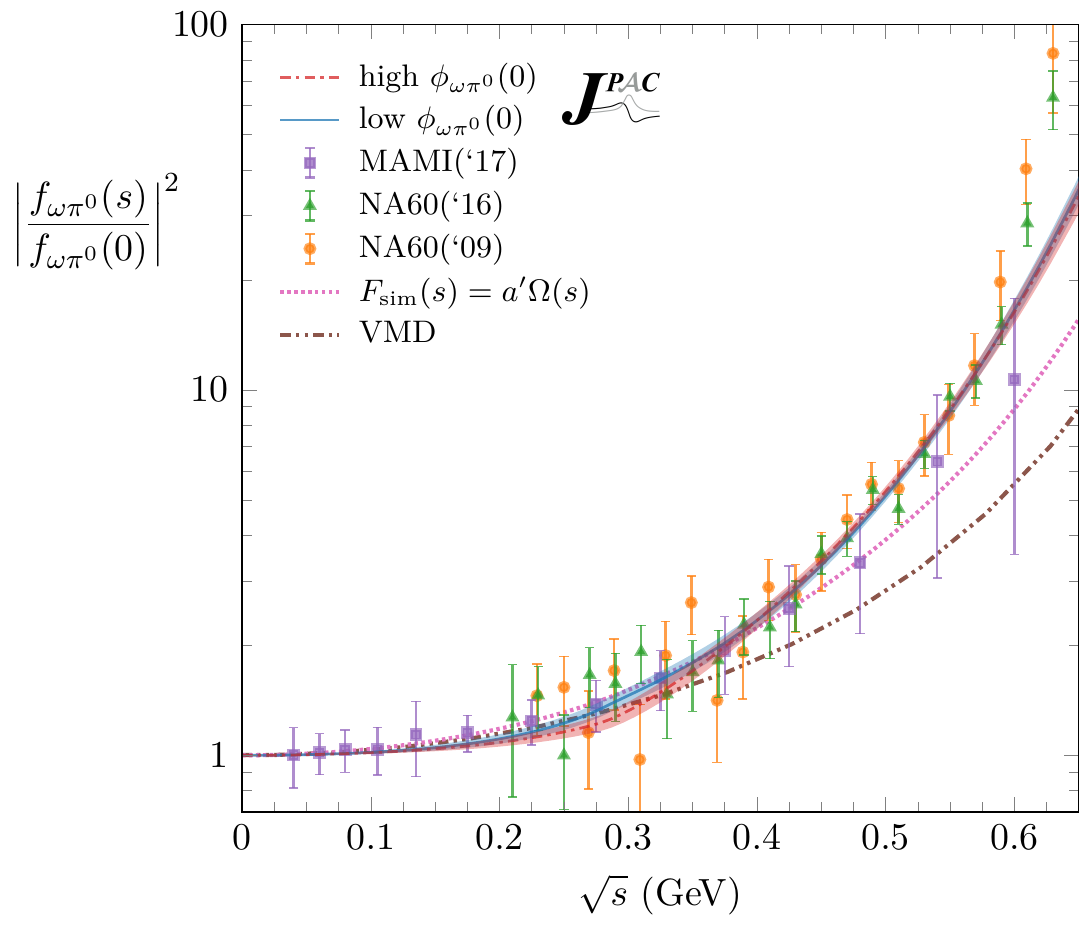}
\caption{Normalized $\omega \pi^0$ TFF, $| f_{\omega\pi^0}(s)/f_{\omega\pi^0}(0)|^2$. The data are taken from Refs.~\cite{Arnaldi:2009aa,Arnaldi:2016pzu,Adlarson:2016hpp}. The lines, and their associated error bands, represent our two different solutions, which overlap almost completely in the $\omega\pi^0$ invariant mass range shown. The case shown here is that corresponding to the ``3 par.'' fit. The curves for the ``2 par.'' case are very similar. For comparison, we also show the Vector Meson Dominance (dot-dot-dashed brown) prediction, and that of the model without KT equations (dotted pink curve) discussed in Subsec.~\ref{subsec:otherapproaches}, \cf Eqs.~\eqref{eqs:simpleOmega}.\label{fig:fit_fop2}}
\end{figure}

The results for the TFF are shown in Fig.~\ref{fig:fit_fop2} for the low and high $\phaf$ solutions. It can be seen that both of them agree very well with the experimental points, except for the highest two points of the NA60 data.\footnote{These two points give a contribution of around $17$ to $\chi^2_\text{NA60}$. However, we note that fits without these two points give similar results as the ones discussed in the text.} Also, it should be noted 
 that both solutions are almost indistinguishable. The largest difference is  at the  $\omega\pi^0$ invariant mass $\sqrt{s} \simeq 0.3\ \text{GeV}$, which is near the $2\pi$ threshold, but even there they are compatible at $1\sigma$ level. Although we will later on compare in detail our results with other approaches, it is worth pointing out here that our theoretical description of the data represents an improvement over previous theoretical analyses~\cite{Terschlusen:2012xw, Danilkin:2014cra, Schneider:2012ez}.

\begin{figure}\centering
\includegraphics{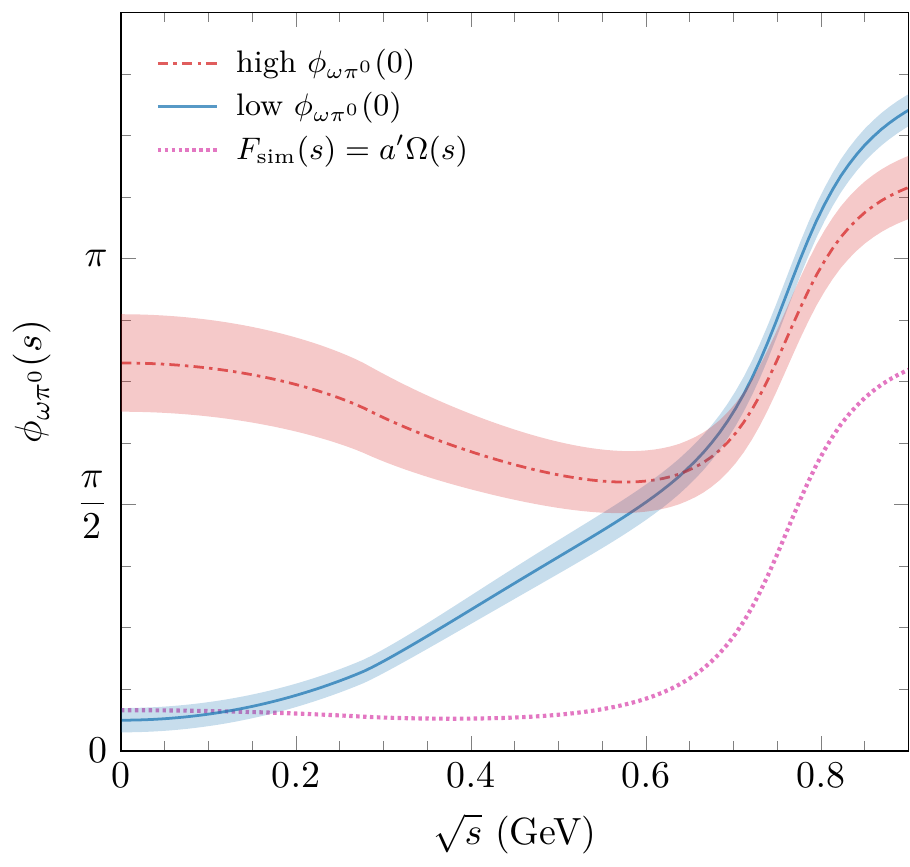}
\caption{Dependence on $s$ of the phase of the $\omega\pi^0$ transition form factor, $\phi_{\omega\pi^0}(s)$, for the two different solutions described in the text. At $s=0$, the phase is given by the fitted parameter $\phi_{\omega\pi^0}(0)$ shown in Table~\ref{table:parameters}. The error bands represent our (correlated) $1\sigma$ uncertainties in the fitted parameters, obtained from a MC analysis of the data. We show here the curves for the ``3 par.'' fit. The phases for the ``2 par.'' case are very similar. For comparison, we also show the prediction of the model without KT equations (dotted pink curve) discussed in Subsec.~\ref{subsec:otherapproaches}, \cf Eqs.~\eqref{eqs:simpleOmega}.\label{fig:fit_fop2_phase}}
\end{figure}

We note that the different phase $\phaf$ in both solutions translates into a difference in the phase of the TFF in a large region of $\omega\pi^0$ invariant mass, up to $\sqrt{s} \simeq 0.6\ \text{GeV}$, as shown in Fig.~\ref{fig:fit_fop2_phase}. For energies $\sqrt{s} \gtrsim 0.6\ \text{GeV}$ the phase motion associated with the $\rho$ meson kicks in, and both solutions approximately converge. This phase, or more properly the phase difference $\phaf - \phi_a$ (see Subsec.~\ref{subsec:TFF}) has not been measured, to the best of our knowledge, and thus Fig.~\ref{fig:fit_fop2_phase} constitutes a prediction for it.\footnote{A different prediction is given in Ref.~\cite{Schneider:2012ez}, as discussed later on in Subsec.~\ref{subsec:otherapproaches}.} Finally, we note that the ``low $\phaf$'' solution is rather close to $\phaf =0$, and the ``high $\phaf$'' is close (but less than the previous one) to $\phaf=\pi$. If the amplitudes were computed from a Lagrangian approach with a  stable $\omega$, the couplings in the Lagrangians would be real. Then, one would expect real values for $a$ and $f_{\omega\pi^0}(0)$, and thus their relative phase could only be $0$ or $\pi$. Anyhow, we find that the inclusion of this phase with a value different from $0$ or $\pi$ improves the description of the data, since they are different from zero by approximately  $2\sigma$.

In what relates to the Dalitz plot parameters, we find good agreement between the input taken from BESIII and our results, see Table~\ref{table:DalitzPlot}, which results in the low $\chi^2_\text{DP}$ shown in Table~\ref{table:parameters}, in the four cases considered (low or high $\phaf$, 2 or 3 Dalitz plot parameters). The largest difference  between observables used in our fit for the ``3 par.'' case is found in $\gamma$. The values that we obtain, $\gamma = \left( 19 \pm 5 \right) \cdot 10^{-3}$ and $ \left( 29 \pm 6 \right) \cdot 10^{-3}$ for the ``low $\phaf$'' and ``high $\phaf$'' solutions, respectively, are both compatible with the experimental one used in the fit, $\gamma = \left( 22 \pm 29  \right) \cdot 10^{-3}$. However, our values are found to be better constrained and indicate that this parameter is non-zero at a $\sim\!3\sigma$ level. Interestingly, the two values of $\gamma$ are only marginally compatible and a more precise measurement of the $\omega \to 3\pi$ Dalitz-plot parameters could help in pinning down the correct solution. A similar argument, though less stringent, can be made for $\beta$ in the ``2 par.'' fits. 

\subsection{Comparison with previous approaches}\label{subsec:otherapproaches}

Our results obtained by solving KT equations for the $\omega \to 3\pi$ amplitude, are compared with those from  Refs.~\cite{Niecknig:2012sj,Danilkin:2014cra}  in Table~\ref{table:DalitzPlot}. The difference between these approaches and ours lies in the subtraction that we have performed on the KT dispersion relations, which introduces an additional free parameter, $b$. In Ref.~\cite{Niecknig:2012sj}, an estimation for this parameter is given by enforcing the once-subtracted DR to be equivalent to the unsubtracted DR. This value, $b_\text{sum} \simeq 0.55 e^{0.15i}\ \text{GeV}^{-2}$, Eq.~\eqref{Eq:SumRuleNum}, turns out to be far away from our fitted $b$ (for any of the fits in Table~\ref{table:parameters}), which reaffirms the need of the extra subtraction. Due to this subtraction, and the fits performed in Subsecs.~\ref{subsec:general} and \ref{subsec:globalfits}, our results for the Dalitz-plot parameters are in agreement with those of the BESIII experiment.

The values of the TFF given by the KT approach without the additional subtraction used in our work for the $\omega \to 3\pi$ amplitude lie systematically below the experimental points~\cite{Schneider:2012ez,Danilkin:2014cra}. In Ref.~\cite{Danilkin:2014cra} it was shown that without the extra subtraction a satisfactory result for the TFF can be obtained only if additional terms are retained in the non-dispersive term (see Fig. 8 of that reference). In contrast, as discussed in Subsec.~\ref{subsec:globalfits}, our results for the TFF are in good agreement with the experimental data. In particular, our approach represents a significant improvement in the description of the higher energy points.\footnote{See also Refs.~\cite{Ananthanarayan:2014pta,Caprini:2015wja}, where the authors use KT supplemented by analyticity and unitarity arguments through the method of unitarity bounds.} 

In Table~\ref{table:DalitzPlot} we show the results obtained in Refs.~\cite{Niecknig:2012sj,Danilkin:2014cra} when the crossed channel effects, which are the essential outcome of the KT equations, are ``turned off'' from the isobar $F(s)$. In practical terms, this is achieved by neglecting the contribution of $\hat{F}(s)$ in Eq.~\eqref{eqs:ourFs}, such that $F(s)$ is simply an Omn\`es function times a constant,
\begin{subequations}\label{eqs:simpleOmega}
\begin{equation}\label{eq:simpleOmega_Fs}
    F_\text{sim}(s) = a'\, \Omega(s)~.
\end{equation}
The reduced full amplitude would then read 
\begin{equation}\label{eq:simpleOmega_Fstu}
    F_\text{sim}(s,t,u) = a' \left( \Omega(s) + \Omega(t) + \Omega(u) \right)~,
\end{equation}
The proportionality constant, $a'$ instead of $a$, is chosen to reproduce the $\omega \to 3\pi$ width, $10^{-2} \left\lvert a' \right\rvert = 2.818(18)\ \text{GeV}^{-3}$, but it is a global constant and does not affect the values of the Dalitz plot parameters. Interestingly, as discussed in Sec.~\ref{sec:introduction}, the Dalitz plot parameters obtained in Refs.~\cite{Niecknig:2012sj,Danilkin:2014cra} in this simplified approach appear to be in better agreement with the recent experimental determination by  BESIII~\cite{Ablikim:2018yen} than those obtained with the crossed channel effects included (but no extra subtraction), \cf Table~\ref{table:DalitzPlot}, rows denoted ``w/o KT'' {\it vs.} those denoted ``w KT'', respectively. In sharp contrast, we show in this work that the results we obtain by keeping the crossed channel effects, and with the additional subtraction, reproduce very well the experimental Dalitz-plot parameters, and are consistent with the $\omega \pi^0$ TFF. We first discuss why the determination of the Dalitz-plot parameters is very similar in our approach (subtracted KT) and in the simpler model (no KT, Eqs.~\eqref{eqs:simpleOmega}). Later on, we will compare the results for the TFF.

\begin{figure}\centering
\includegraphics{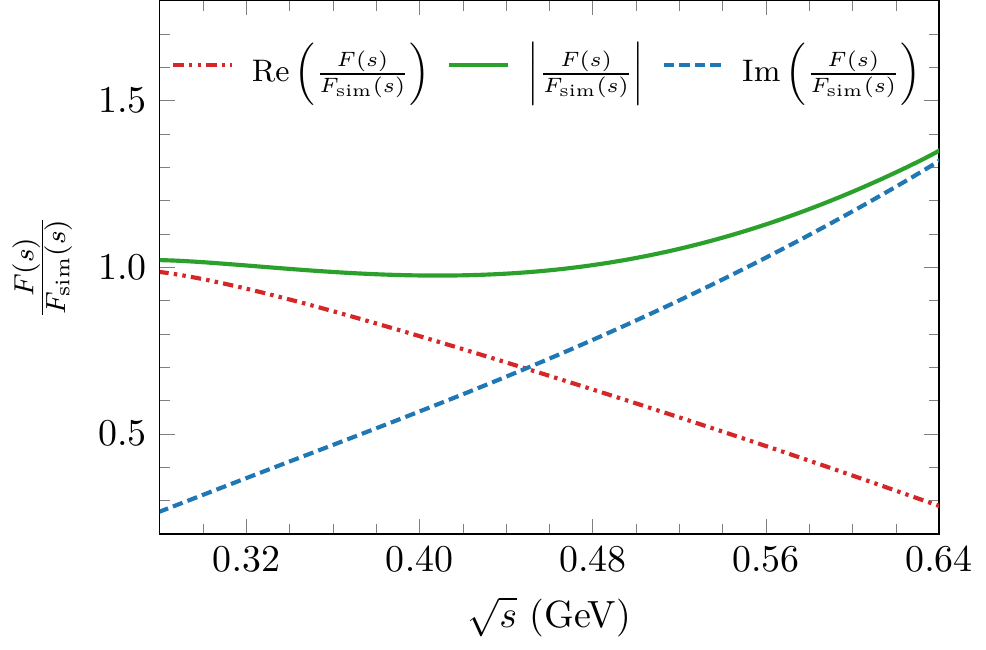}
\caption{Absolute value (green solid), real (red dot-dot-dashed) and imaginary (blue dashed) parts of the $F(s)/F_\text{sim}(s)$ ratio, as described in the text. The ratio is shown in the physical decay range of $s$.\label{fig:cancellation_A}}
\end{figure}

\begin{figure}\centering
\includegraphics{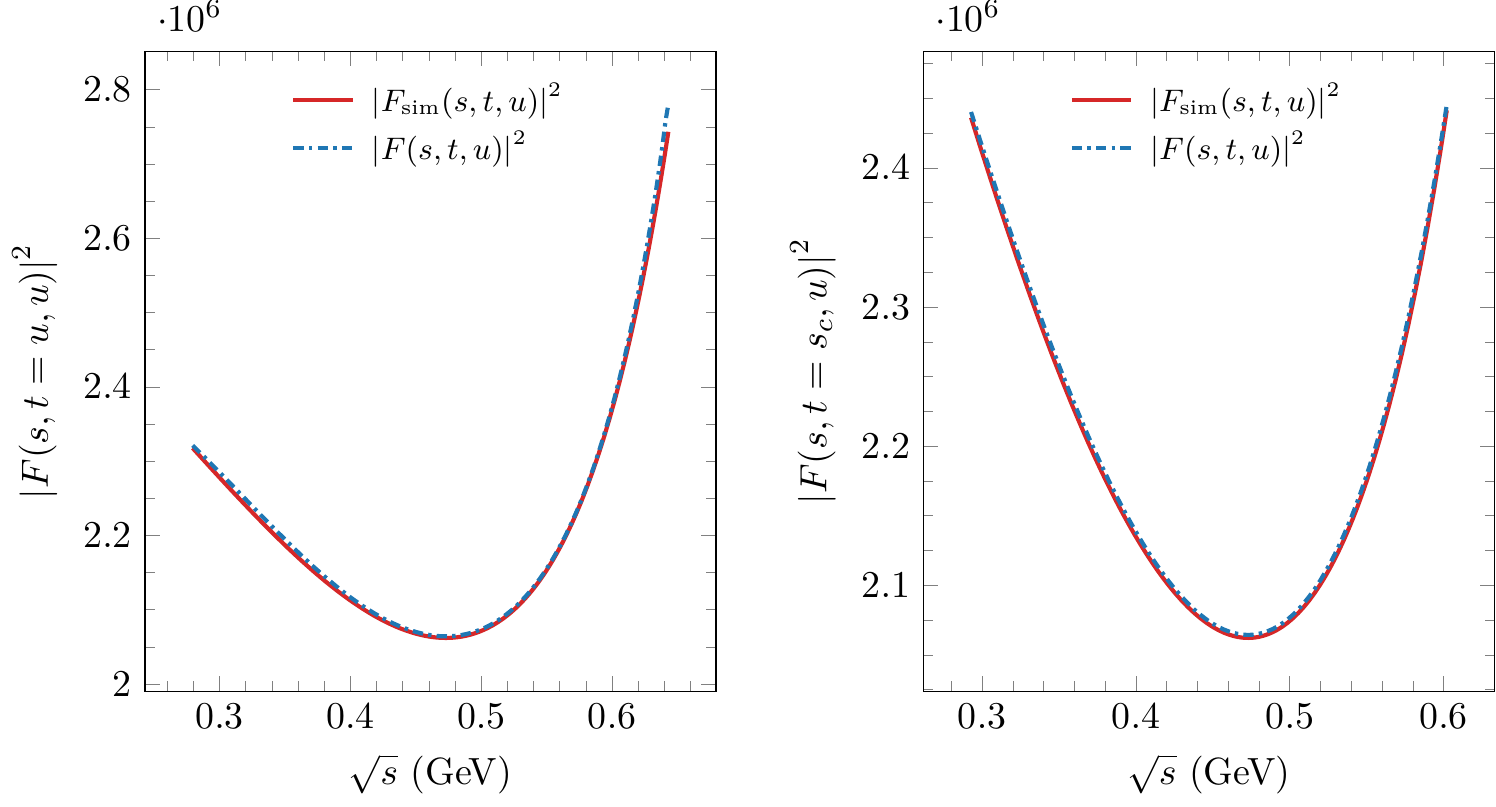}
\caption{Modulus squared of the amplitudes in our full model ($F(s,t,u)$, dashed blue lines) and in the simplified model ($F_\text{sim}(s,t,u)$, Eqs.~\eqref{eqs:simpleOmega}, solid red lines). The left (right) plot shows the functions along the $t=u$ ($t=s_c$) lines of the Mandelstam plane. The ratio is shown in the physical decay range of $s$ for each of the cases considered.
\label{fig:cancellation_B}}
\end{figure}

The aforementioned agreement is clear, as can be seen in Table~\ref{table:DalitzPlot}, and hence there must be some sort of cancellation that ``brings back'' our full subtracted KT approach into the simpler, no KT model. Naively, if one thinks that the KT formalism is overestimating the crossed channel effects, it would be expected that this cancellation would occur in the isobar amplitude itself, $F(s)$, \ie, that the effect of the crossed channels is mostly linear and thus can be  absorbed by the additional subtraction constant, $b$. In this case, the ratio $\displaystyle F(s)/F_\text{sim}(s)$ should be essentially constant. We show in Fig.~\ref{fig:cancellation_A} that this is certainly not the case, although the modulus of the ratio is still around $1$. Here, we are taking the parameters of the ``low $\phaf$'' solution for the ``3 par.'' case, but similar results are obtained in the other fits. This demonstrates that the cancellation is not trivial, as one would expect if the crossed channel effects were simply being overestimated. 

The cancellation must thus occur at the level of the squared amplitude, $\left\lvert F(s,t,u) \right\rvert^2$ $= \left\lvert F(s) + F(t) + F(u) \right\rvert^2$. In Fig.~\ref{fig:cancellation_B} we show $\left\lvert F(s,t,u) \right\rvert^2$ for $\sqrt{s}$ in the physical decay region for two lines across the $(s,t,u)$ plane, namely, $t=u$ and $t=s_c$ (respectively corresponding to $X=0$ and $Y=-\sqrt{3}X$ in the usual $(X,Y)$ Dalitz plot variables, \cf Eq.~\eqref{eq:dalitzxy}). We also show in the figures the function $\left\lvert F_\text{sim}(s,t,u) \right\rvert^2$, \ie, the full amplitude squared for the simpler model [Eqs.~\eqref{eqs:simpleOmega}] discussed above. It can be seen in Fig.~\ref{fig:cancellation_B} that the differences between both squared moduli are quite small. We also find that the phase difference between our $F(s,t,u)$ and $F_\text{sim}(s,t,u)$ is essentially constant. This large cancellation explains the coincidence of the results for the Dalitz plot parameters in both approaches.\footnote{The cancellation is also aided by the fact that the nominal $\rho$-meson mass lies outside the physical $\omega \to 3\pi$ decay region, and hence the Omn\`es function is still relatively smooth.}

As a result of the above discussion, one might question the necessity of the full approach if, after all, the rather simpler description with no subtractions and no crossed channel effects, Eq.~\eqref{eqs:simpleOmega}, seems to work just fine. However, it must be noted that this simpler model only describes well the $\omega \to 3\pi$ Dalitz-plot parameters, but not the distribution for the $\phi \to 3\pi$ decay \cite{Niecknig:2012sj} nor the more precise experimental information on the $\omega\pi^0$ TFF. In Ref.~\cite{Schneider:2012ez} it is shown that a model which ignores the crossed-channel effects by inserting $f_1(s) = a' \Omega(s)$ into Eq.~\eqref{Eq:OmegaPiFF1sub} gives a result well below the experimental points (see Fig.~5 of Ref.~\cite{Schneider:2012ez}). We could also take the partial wave that results from Eqs.~\eqref{eq:simpleOmega_Fs} and \eqref{eq:simpleOmega_Fstu}, which is given by
\begin{equation}\label{eq:simpleOmega_f1}
f_{1,\text{sim}}(s) = a' \left( \Omega(s) + 3\int_{-1}^{1}\frac{dz_{s}}{2} \; (1-z_{s}^{2}) \; \Omega(t(s,z_{s})) \right)~.
\end{equation}\end{subequations}
This model, when introduced into Eq.~\eqref{Eq:OmegaPiFF1sub}, produces the result shown as a pink dotted line in Fig.~\ref{fig:fit_fop2}, which is well below the experimental points and our results.\footnote{The phase of the TFF predicted in this case is also quite different from our results, see Fig.~\ref{fig:fit_fop2_phase}.} This result for the TFF is very similar to that of Ref.~\cite{Schneider:2012ez} mentioned above.

In summary, from a phenomenological point of view, our description of the Dalitz-plot parameters and of the TFF using a once-subtracted version of the KT equations is (not surprisingly) better than that obtained with unsubtracted KT equations \cite{Niecknig:2012sj,Schneider:2012ez,Danilkin:2014cra}. On the other hand, the simpler model of Eqs.~\eqref{eqs:simpleOmega}, in which the KT effects are ignored, describe properly the Dalitz-plot parameters (see the discussion above about Figs.~\ref{fig:cancellation_A} and \ref{fig:cancellation_B}), but not the TFF data. Therefore, it seems that our approach, in which a KT equation for the $\omega \to 3\pi$ amplitude is solved with an additional subtraction, is the minimal theoretical setup that is able to simultaneously describe both sets of data. From a more theoretical perspective, it is clear that the crossed channel effects must be present in any $2 \to 2$ or $1 \to 3$ amplitude, even if they are negligible or can be mimicked by polynomial terms~\cite{Albaladejo:2018gif}. The KT formalism offers a simple framework which allows to provide the partial waves in the direct channel with left hand cuts in terms of the isobars of the crossed channels, while allowing to incorporate crossing symmetry, unitarity and, to some extent,\footnote{For discussion on this topic, see \eg Ref.~\cite{Albaladejo:2019huw,Oller:2019opk,Oller:2020guq} and references therein.} analyticity.

\section{Outlook}\label{sec:conclusions}

\paragraph{Summary.-} In this work we have explored the benefits of a simultaneous analysis of the $\omega\to3\pi$ decay and the $\omega\pi^0$  transition form factor. The motivation for this study is manifold. First, from the point of view of strong interactions, the decay $\omega\to3\pi$ offers a good environment to study the dynamics of the $\pi\pi$ subsystems under rather clean conditions. Second, the BESIII collaboration has reported a high-statistics measurement of the $\omega\to3\pi$ Dalitz plot distribution, and pointed out a possible overestimation of the crossed-channel contributions in the KT equations. Third, there are recent data on the shape of the $\omega\pi^{0}$ TFF from the MAMI and NA60 collaborations making such an analysis of timely interest.

For the $\omega\to3\pi$ amplitude we follow a dispersive representation with subtractions that emerges from the solution of the KT equation\,\cite{Niecknig:2012sj,Danilkin:2014cra}.  It thus satisfies the constraints posed by analyticity (to some extent), crossing symmetry and (elastic) unitarity, and it is completely determined by the $\pi\pi$ $P$-wave scattering phase shift, except for the values of the subtraction constants. In this work we have performed one subtraction, which introduces an additional free parameter, $b$, apart from the usual global normalization $a$ that is fixed from the partial decay width. We fix this extra parameter, which is characterized by its modulus $|b|$ and phase $\phi_{b}$, from fits to experimental data. The $\omega\to3\pi$ amplitude, in turn, enters the once-subtracted dispersive parametrization of the $\omega\pi^{0}$ TFF Eq.\,(\ref{Eq:OmegaPiFF1sub}), introducing its phase at $s=0$, $\phaf$, as a new ingredient of this work.

Our first analysis proceeds in two steps.
On a first step, we use the two different sets of Dalitz-plot parameters given by BESIII and the corresponding partial decay widths to fix all free parameters $(|a|,|b|,\phi_{b},|f_{\omega\pi^{0}}(0)|)$ except for $\phi_{\omega\pi^{0}}(0)$. These results bring us to a first relevant observation: the value of the subtraction constant $b$ needed to faithfully reproduce the Dalitz-plot parameters is found to be significantly different (see Fig.\,\ref{fig:sumrule}) from the sum-rule value estimated from the unsubtracted version of the KT equations. On a second step, the dependence of the $\omega\pi^{0}$ TFF on $\phi_{\omega\pi^{0}}(0)$ is studied in relation to the MAMI and NA60 data. It is found that there are two well separated minima in this variable.

We have also performed a combined analysis to all available experimental information including Dalitz-plot parameters and form-factor data, and observed that the two solutions for $\phaf$ remain. Interestingly enough, the values for the subtraction constant $b$ obtained from the joint fits have a much better constrained uncertainty than that in the individual fits to the BESIII Dalitz-plot parameters (see Fig.~\ref{fig:sumrule}), however being in perfect agreement with it. This reaffirms the need of the additional subtraction constant.

From the Dalitz-plot parameters associated to our combined fits (see Table \ref{table:DalitzPlot}), we can draw a second relevant observation. 
While the values that we obtain for the Dalitz-plot parameters are found to be in agreement with the experimental ones, our values carry a smaller error and indicate a statistical significance for the the Dalitz-plot parameter $\gamma$ of $\sim\!3\sigma$. 
Furthermore, our results for the normalized $\omega\pi^{0}$ TFF (Fig.~\ref{fig:fit_fop2}) show a satisfactory description of the experimental data, except for the highest two points of the NA60 collaboration. 

\paragraph{Open questions.-} Even though we achieved a simultaneous description of the Dalitz-plot parameters and the TFF data, it comes as a surprise that the predictions for the $\omega \to 3\pi$ amplitude are so different between the unsubtracted and once-subtracted versions of the KT equations. (This can be visualized either in the discrepancy between the Dalitz-plot parameters in both cases, or in the large difference between the fitted subtraction constant $b$ respect to the sum-rule expectation.) Moreover, this does not seem to happen in $\phi\to 3\pi$, despite the larger phase space, which makes this difference even more intriguing. 

It is also important to note that, due to the goal of our work, the analysis of the $\omega\pi^0$ TFF has been restricted to the relatively low energy region of the NA60 ($\omega \to \pi^0 \mu^+ \mu^-$) and MAMI ($\omega \to \pi^0 e^+ e^-$) data. Because of this, we have not explored the higher energy region beyond the $\omega\pi^{0}$ threshold, where there are experimental data~\cite{Akhmetshin:2003ag,Achasov:2012zza,Achasov:2013btb,Achasov:2016zvn} coming from the reactions $e^+ e^- \to \omega \pi^0$. To do so would require to consider also higher resonances in the $\pi\pi$ phase shifts, something clearly outside the scope of the present analysis. Furthermore, the NA60 data currently have much smaller uncertainties than the MAMI ones, which translates into the fact that our fits to the TFF have been dominated by the former, with almost no influence of the latter. The NA60 data drive the TFF curve towards higher values (even more if one aims to describe also the last two NA60 data points), which can certainly impact the extrapolation to higher energies. 

Therefore, we hope that our study strengthens the case for a reanalysis of all these decays and/or new measurements thereof, either to reduce uncertainties or to address eventual incompatibilities.

%***********************************************
\acknowledgments
This work was supported by the U.S.~Department of Energy under Grants
No.~DE-AC05-06OR23177 % jlab
and No.~DE-FG02-87ER40365, % IU grant
the U.S.~National Science Foundation under Grant 
No.~PHY-1415459. % Adam, PIF
The work of I.D. was supported by the Deutsche Forschungsgemeinschaft (DFG, German Research Foundation), in part through the Collaborative Research Center [The Low-Energy Frontier of the Standard Model, Projektnummer 204404729 - SFB 1044], and in part through the Cluster of Excellence [Precision Physics, Fundamental Interactions, and Structure of Matter] (PRISMA$^+$ EXC 2118/1) within the German Excellence Strategy (Project ID 39083149).
The work of S.GS has been supported in part by the National Science Foundation (PHY-1714253).
%and by the U.S. Department of Energy under Grant No.\,DE-FG02-87ER40365.
%by the Ministerio de Ciencia, Innovaci\'on y Universidades (Spain) under Grants No.~FPA2016-77313-P and No.~FPA2016-75654-C2-2-P, %% Miguel, national projects
%by  PAPIIT-DGAPA (UNAM, Mexico) under Grant No.~IA101819, % Cesar 
%and CONACYT (Mexico) under Grants No.~251817,  %Cesar
%No.~734789 %Jorge
%and No.~A1-S-21389, %Cesar
%by the DFG [Projektnummer 204404729 - SFB 1044] and in part through the Cluster of Excellence (PRISMA+ EXC
%2118/1) within the German Excellence Strategy (Project ID 39083149). %Igor
V.M. is supported by the Comunidad Aut\'onoma de Madrid through the Programa de Atracci\'on de Talento Investigador 2018 (Modalidad 1). %Vincent
The work of C.F.-R. is supported by PAPIIT-DGAPA (UNAM, Mexico) under Grant No. IA101819 and by CONACYT (Mexico) under Grant No. A1-S-21389. % Cesar

\appendix
\section{Statistical analysis}\label{sec:statistics}

In this Appendix, we give some details about the MC statistical analysis performed in Subsec.~\ref{subsec:globalfits} for the global fits. For each of the four fits considered in Table~\ref{table:parameters}, we generate $\mathcal{O}(10^4)$ sets of the data (resampling) described in Subsec.~\ref{subsec:general}, each single datum following a gaussian distribution. For each of these sets, a fit is performed and each of the output quantities of our work (DP parameters, TFF, etc.) are computed for that fit. In this way, all possible known correlations are taken into account. The values obtained in this work quoted in Tables~\ref{table:DalitzPlot} and \ref{table:parameters}, as well as those represented in Figs.~\ref{fig:fig_reim_fs}, \ref{fig:fit_fop2}, and \ref{fig:fit_fop2_phase} are the average value and the standard deviation of those quantities in all the fits generated.

In the histograms of Fig.~\ref{fig:histogramsDalitzPar} we show the probability distribution of the fitted parameters obtained in our MC analysis for both the low and high $\phaf$ solutions. We show the ``3 par.'' case, but similar results are seen for the ``2 par.'' case. In general, the parameters are seen to follow a Gaussian distribution, although some deviations are seen from this behaviour, specially for $\modb$ and $\phab$. This non-gaussianity is, of course, inherited from the $\chi^2_\text{DP}$ function, as can be seen in Fig.~\ref{fig:sumrule}.

The correlation parameter between the fitted parameters and/or the computed quantities can be calculated in a standard way. However, the two-dimensional distributions are not always Gaussian, and we therefore prefer to show the two-dimensional projections of (a small sample of) our MC simulations in Fig.~\ref{fig:corrMC}.

\begin{figure}\centering
\includegraphics{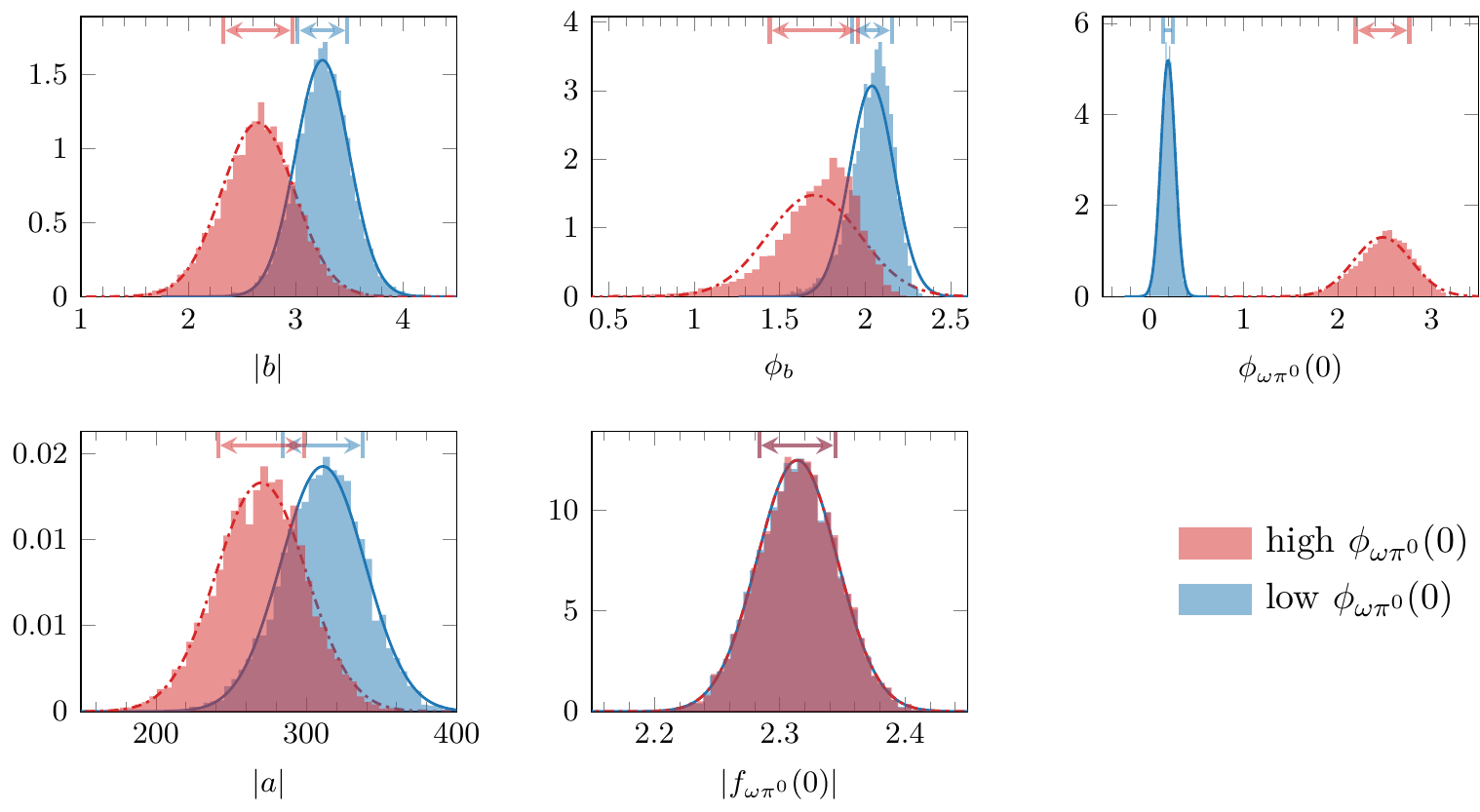}\\
\caption{One dimensional distributions of the free parameters in the Monte Carlo analysis performed, for the ``3 par.'' case fit. A gaussian distribution with the average and error quoted in Table~\ref{table:parameters} is superimposed for each solution. The double arrows on the upper part of each histogram represent the $1\sigma$ uncertainty intervals.\label{fig:histogramsFreePar}}
\end{figure}

\begin{figure}\centering
\includegraphics{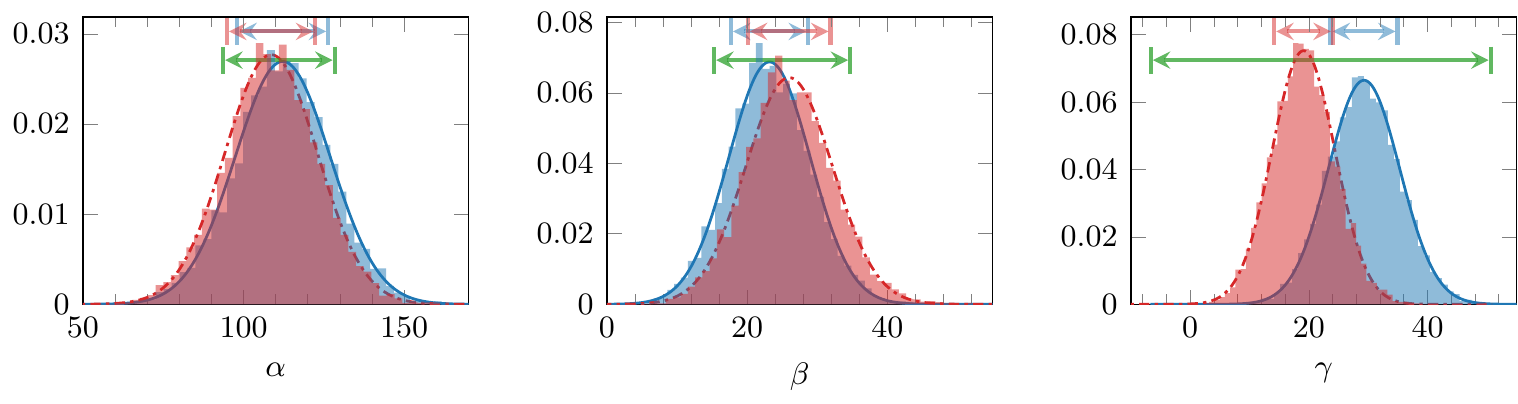}\\
\includegraphics{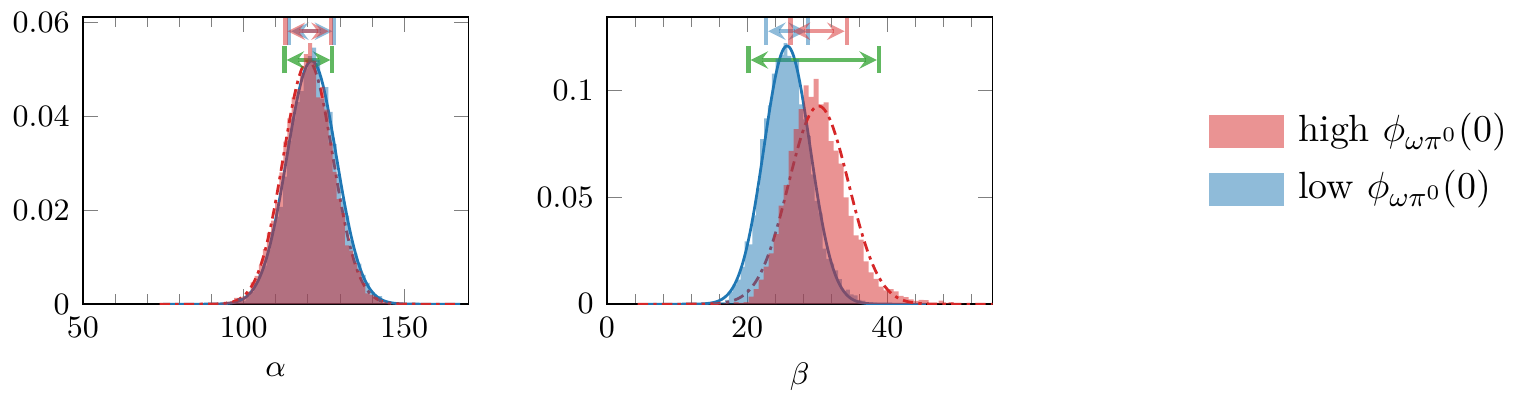}
\caption{One dimensional distributions of our computed Dalitz plot parameters in the Monte Carlo analysis performed, for the ``3 par.'' case (top row) and ``2 par.'' case (bottom row). A Gaussian distribution with the average and error quoted in Table~\ref{table:DalitzPlot} is superimposed for each solution. The double arrows on the upper part of each histogram represent the $1\sigma$ uncertainty intervals. The green line represents the BESIII experimental determination, also shown in Table~\ref{table:DalitzPlot}. \label{fig:histogramsDalitzPar}}
\end{figure}

\begin{figure}\centering
\includegraphics{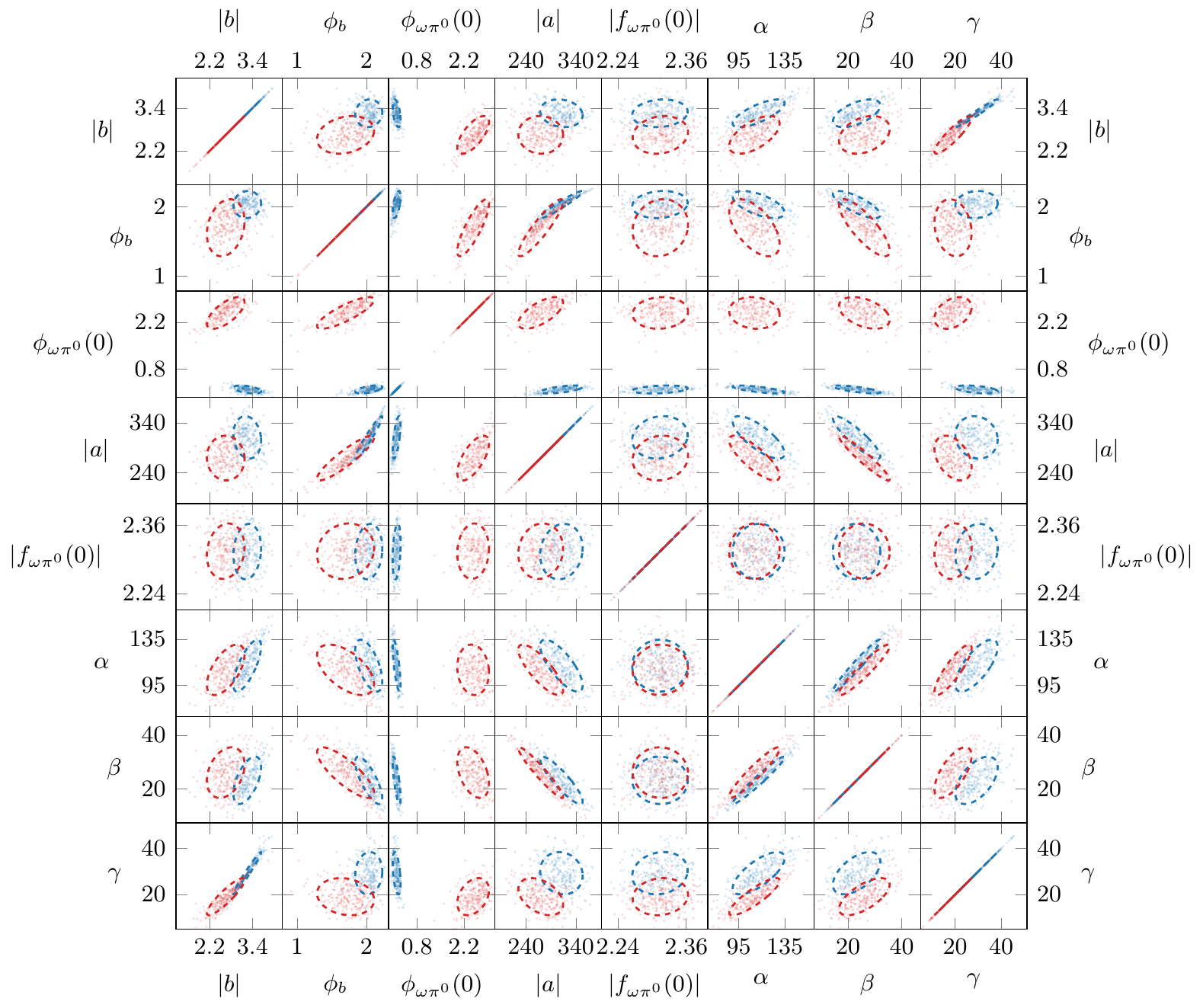}
\caption{Two dimensional projections of the MC simulations, for the fitted and Dalitz-plot parameters. The blue (red) small points represent a pair of values obtained in a single MC simulation for the low (high) $\phaf$ solution. The ellipses represent the two-parameter $1\sigma$ contours computed with the average and standard deviation, and correlation parameter of each pair. This provides a pictorial representation of the correlation and accounts for the non-gaussianity of the distributions.\label{fig:corrMC}}
\end{figure}

%***********************************************

%\newpage

%\bibliographystyle{apsrev4-1_MOD}
\bibliographystyle{apsrev4-2_MOD}
\bibliography{refs}

\begin{thebibliography}{80}%
\makeatletter
\providecommand \@ifxundefined [1]{%
 \@ifx{#1\undefined}
}%
\providecommand \@ifnum [1]{%
 \ifnum #1\expandafter \@firstoftwo
 \else \expandafter \@secondoftwo
 \fi
}%
\providecommand \@ifx [1]{%
 \ifx #1\expandafter \@firstoftwo
 \else \expandafter \@secondoftwo
 \fi
}%
\providecommand \natexlab [1]{#1}%
\providecommand \enquote  [1]{``#1''}%
\providecommand \bibnamefont  [1]{#1}%
\providecommand \bibfnamefont [1]{#1}%
\providecommand \citenamefont [1]{#1}%
\providecommand \href@noop [0]{\@secondoftwo}%
\providecommand \href [0]{\begingroup \@sanitize@url \@href}%
\providecommand \@href[1]{\@@startlink{#1}\@@href}%
\providecommand \@@href[1]{\endgroup#1\@@endlink}%
\providecommand \@sanitize@url [0]{\catcode `\\12\catcode `\$12\catcode
  `\&12\catcode `\#12\catcode `\^12\catcode `\_12\catcode `\%12\relax}%
\providecommand \@@startlink[1]{}%
\providecommand \@@endlink[0]{}%
\providecommand \url  [0]{\begingroup\@sanitize@url \@url }%
\providecommand \@url [1]{\endgroup\@href {#1}{\urlprefix }}%
\providecommand \urlprefix  [0]{URL }%
\providecommand \Eprint [0]{\href }%
\providecommand \doibase [0]{https://doi.org/}%
\providecommand \selectlanguage [0]{\@gobble}%
\providecommand \bibinfo  [0]{\@secondoftwo}%
\providecommand \bibfield  [0]{\@secondoftwo}%
\providecommand \translation [1]{[#1]}%
\providecommand \BibitemOpen [0]{}%
\providecommand \bibitemStop [0]{}%
\providecommand \bibitemNoStop [0]{.\EOS\space}%
\providecommand \EOS [0]{\spacefactor3000\relax}%
\providecommand \BibitemShut  [1]{\csname bibitem#1\endcsname}%
\let\auto@bib@innerbib\@empty
%</preamble>
\bibitem [{\citenamefont {Aaij}\ \emph {et~al.}(2014)\citenamefont {Aaij} \emph
  {et~al.}}]{Aaij:2014jqa}%
  \BibitemOpen
  \bibfield  {author} {\bibinfo {author} {\bibfnamefont {R.}~\bibnamefont
  {Aaij}} \emph {et~al.} (\bibinfo {collaboration} {LHCb}),\ }\href
  {https://doi.org/10.1103/PhysRevLett.112.222002} {\bibfield  {journal}
  {\bibinfo  {journal} {Phys. Rev. Lett.}\ }\textbf {\bibinfo {volume} {112}},\
  \bibinfo {pages} {222002} (\bibinfo {year} {2014})},\ \Eprint
  {https://arxiv.org/abs/1404.1903} {arXiv:1404.1903 [hep-ex]}\BibitemShut
  {NoStop}%
%%CITATION = ARXIV:1404.1903;%%
\bibitem [{\citenamefont {Aaij}\ \emph {et~al.}(2015)\citenamefont {Aaij} \emph
  {et~al.}}]{Aaij:2015tga}%
  \BibitemOpen
  \bibfield  {author} {\bibinfo {author} {\bibfnamefont {R.}~\bibnamefont
  {Aaij}} \emph {et~al.} (\bibinfo {collaboration} {LHCb}),\ }\href
  {https://doi.org/10.1103/PhysRevLett.115.072001} {\bibfield  {journal}
  {\bibinfo  {journal} {Phys. Rev. Lett.}\ }\textbf {\bibinfo {volume} {115}},\
  \bibinfo {pages} {072001} (\bibinfo {year} {2015})},\ \Eprint
  {https://arxiv.org/abs/1507.03414} {arXiv:1507.03414 [hep-ex]}\BibitemShut
  {NoStop}%
%%CITATION = ARXIV:1507.03414;%%
\bibitem [{\citenamefont {Al~Ghoul}\ \emph {et~al.}(2016)\citenamefont
  {Al~Ghoul} \emph {et~al.}}]{Ghoul:2015ifw}%
  \BibitemOpen
  \bibfield  {author} {\bibinfo {author} {\bibfnamefont {H.}~\bibnamefont
  {Al~Ghoul}} \emph {et~al.} (\bibinfo {collaboration} {GlueX}),\ }\href
  {https://doi.org/10.1063/1.4949369} {\bibfield  {journal} {\bibinfo
  {journal} {AIP Conf. Proc.}\ }\textbf {\bibinfo {volume} {1735}},\ \bibinfo
  {pages} {020001} (\bibinfo {year} {2016})},\ \Eprint
  {https://arxiv.org/abs/1512.03699} {arXiv:1512.03699 [nucl-ex]}\BibitemShut
  {NoStop}%
\bibitem [{\citenamefont {Krinner}(2018)}]{Krinner:2017vch}%
  \BibitemOpen
  \bibfield  {author} {\bibinfo {author} {\bibfnamefont {F.~M.}\ \bibnamefont
  {Krinner}} (\bibinfo {collaboration} {COMPASS}),\ }\href
  {https://doi.org/10.22323/1.310.0034} {\bibfield  {journal} {\bibinfo
  {journal} {PoS}\ }\textbf {\bibinfo {volume} {Hadron2017}},\ \bibinfo {pages}
  {034} (\bibinfo {year} {2018})},\ \Eprint {https://arxiv.org/abs/1711.10828}
  {arXiv:1711.10828 [hep-ph]}\BibitemShut {NoStop}%
\bibitem [{\citenamefont {Aghasyan}\ \emph {et~al.}(2018)\citenamefont
  {Aghasyan} \emph {et~al.}}]{Akhunzyanov:2018pnr}%
  \BibitemOpen
  \bibfield  {author} {\bibinfo {author} {\bibfnamefont {M.}~\bibnamefont
  {Aghasyan}} \emph {et~al.} (\bibinfo {collaboration} {COMPASS}),\ }\href
  {https://doi.org/10.1103/PhysRevD.98.092003} {\bibfield  {journal} {\bibinfo
  {journal} {Phys. Rev. D}\ }\textbf {\bibinfo {volume} {98}},\ \bibinfo
  {pages} {092003} (\bibinfo {year} {2018})},\ \Eprint
  {https://arxiv.org/abs/1802.05913} {arXiv:1802.05913 [hep-ex]}\BibitemShut
  {NoStop}%
\bibitem [{\citenamefont {Ablikim}\ \emph {et~al.}(2020)\citenamefont {Ablikim}
  \emph {et~al.}}]{Ablikim:2019hff}%
  \BibitemOpen
  \bibfield  {author} {\bibinfo {author} {\bibfnamefont {M.}~\bibnamefont
  {Ablikim}} \emph {et~al.},\ }\href
  {https://doi.org/10.1088/1674-1137/44/4/040001} {\bibfield  {journal}
  {\bibinfo  {journal} {Chin. Phys. C}\ }\textbf {\bibinfo {volume} {44}},\
  \bibinfo {pages} {040001} (\bibinfo {year} {2020})},\ \Eprint
  {https://arxiv.org/abs/1912.05983} {arXiv:1912.05983 [hep-ex]}\BibitemShut
  {NoStop}%
\bibitem [{\citenamefont {Mokeev}\ \emph {et~al.}(2020)\citenamefont {Mokeev}
  \emph {et~al.}}]{Mokeev:2020hhu}%
  \BibitemOpen
  \bibfield  {author} {\bibinfo {author} {\bibfnamefont {V.}~\bibnamefont
  {Mokeev}} \emph {et~al.},\ }\href
  {https://doi.org/10.1016/j.physletb.2020.135457} {\bibfield  {journal}
  {\bibinfo  {journal} {Phys. Lett. B}\ }\textbf {\bibinfo {volume} {805}},\
  \bibinfo {pages} {135457} (\bibinfo {year} {2020})},\ \Eprint
  {https://arxiv.org/abs/2004.13531} {arXiv:2004.13531 [nucl-ex]}\BibitemShut
  {NoStop}%
\bibitem [{\citenamefont {Brice\~no}\ \emph {et~al.}(2018)\citenamefont
  {Brice\~no}, \citenamefont {Dudek},\ and\ \citenamefont
  {Young}}]{Briceno:2017max}%
  \BibitemOpen
  \bibfield  {author} {\bibinfo {author} {\bibfnamefont {R.~A.}\ \bibnamefont
  {Brice\~no}}, \bibinfo {author} {\bibfnamefont {J.~J.}\ \bibnamefont
  {Dudek}},\ and\ \bibinfo {author} {\bibfnamefont {R.~D.}\ \bibnamefont
  {Young}},\ }\href {https://doi.org/10.1103/RevModPhys.90.025001} {\bibfield
  {journal} {\bibinfo  {journal} {Rev. Mod. Phys.}\ }\textbf {\bibinfo {volume}
  {90}},\ \bibinfo {pages} {025001} (\bibinfo {year} {2018})},\ \Eprint
  {https://arxiv.org/abs/1706.06223} {arXiv:1706.06223 [hep-lat]}\BibitemShut
  {NoStop}%
%%CITATION = ARXIV:1706.06223;%%
\bibitem [{\citenamefont {Jackura}\ \emph {et~al.}(2019)\citenamefont
  {Jackura}, \citenamefont {Dawid}, \citenamefont {Fernández-Ram\'irez},
  \citenamefont {Mathieu}, \citenamefont {Mikhasenko}, \citenamefont {Pilloni},
  \citenamefont {Sharpe},\ and\ \citenamefont {Szczepaniak}}]{Jackura:2019bmu}%
  \BibitemOpen
  \bibfield  {author} {\bibinfo {author} {\bibfnamefont {A.~W.}\ \bibnamefont
  {Jackura}}, \bibinfo {author} {\bibfnamefont {S.~M.}\ \bibnamefont {Dawid}},
  \bibinfo {author} {\bibfnamefont {C.}~\bibnamefont {Fernández-Ram\'irez}},
  \bibinfo {author} {\bibfnamefont {V.}~\bibnamefont {Mathieu}}, \bibinfo
  {author} {\bibfnamefont {M.}~\bibnamefont {Mikhasenko}}, \bibinfo {author}
  {\bibfnamefont {A.}~\bibnamefont {Pilloni}}, \bibinfo {author} {\bibfnamefont
  {S.~R.}\ \bibnamefont {Sharpe}},\ and\ \bibinfo {author} {\bibfnamefont
  {A.~P.}\ \bibnamefont {Szczepaniak}},\ }\href
  {https://doi.org/10.1103/PhysRevD.100.034508} {\bibfield  {journal} {\bibinfo
   {journal} {Phys. Rev.}\ }\textbf {\bibinfo {volume} {D100}},\ \bibinfo
  {pages} {034508} (\bibinfo {year} {2019})},\ \Eprint
  {https://arxiv.org/abs/1905.12007} {arXiv:1905.12007 [hep-ph]}\BibitemShut
  {NoStop}%
%%CITATION = ARXIV:1905.12007;%%
\bibitem [{\citenamefont {Brice\~no}\ \emph {et~al.}(2019)\citenamefont
  {Brice\~no}, \citenamefont {Hansen}, \citenamefont {Sharpe},\ and\
  \citenamefont {Szczepaniak}}]{Briceno:2019muc}%
  \BibitemOpen
  \bibfield  {author} {\bibinfo {author} {\bibfnamefont {R.~A.}\ \bibnamefont
  {Brice\~no}}, \bibinfo {author} {\bibfnamefont {M.~T.}\ \bibnamefont
  {Hansen}}, \bibinfo {author} {\bibfnamefont {S.~R.}\ \bibnamefont {Sharpe}},\
  and\ \bibinfo {author} {\bibfnamefont {A.~P.}\ \bibnamefont {Szczepaniak}},\
  }\href {https://doi.org/10.1103/PhysRevD.100.054508} {\bibfield  {journal}
  {\bibinfo  {journal} {Phys. Rev.}\ }\textbf {\bibinfo {volume} {D100}},\
  \bibinfo {pages} {054508} (\bibinfo {year} {2019})},\ \Eprint
  {https://arxiv.org/abs/1905.11188} {arXiv:1905.11188 [hep-lat]}\BibitemShut
  {NoStop}%
%%CITATION = ARXIV:1905.11188;%%
\bibitem [{\citenamefont {Mai}\ \emph {et~al.}(2019)\citenamefont {Mai},
  \citenamefont {Culver}, \citenamefont {Alexandru}, \citenamefont
  {D{\"o}ring},\ and\ \citenamefont {Lee}}]{Mai:2019pqr}%
  \BibitemOpen
  \bibfield  {author} {\bibinfo {author} {\bibfnamefont {M.}~\bibnamefont
  {Mai}}, \bibinfo {author} {\bibfnamefont {C.}~\bibnamefont {Culver}},
  \bibinfo {author} {\bibfnamefont {A.}~\bibnamefont {Alexandru}}, \bibinfo
  {author} {\bibfnamefont {M.}~\bibnamefont {D{\"o}ring}},\ and\ \bibinfo
  {author} {\bibfnamefont {F.~X.}\ \bibnamefont {Lee}},\ }\href
  {https://doi.org/10.1103/PhysRevD.100.114514} {\bibfield  {journal} {\bibinfo
   {journal} {Phys. Rev.}\ }\textbf {\bibinfo {volume} {D100}},\ \bibinfo
  {pages} {114514} (\bibinfo {year} {2019})},\ \Eprint
  {https://arxiv.org/abs/1908.01847} {arXiv:1908.01847 [hep-lat]}\BibitemShut
  {NoStop}%
%%CITATION = ARXIV:1908.01847;%%
\bibitem [{\citenamefont {Culver}\ \emph {et~al.}(2019)\citenamefont {Culver},
  \citenamefont {Mai}, \citenamefont {Brett}, \citenamefont {Alexandru},\ and\
  \citenamefont {D{\"o}ring}}]{Culver:2019vvu}%
  \BibitemOpen
  \bibfield  {author} {\bibinfo {author} {\bibfnamefont {C.}~\bibnamefont
  {Culver}}, \bibinfo {author} {\bibfnamefont {M.}~\bibnamefont {Mai}},
  \bibinfo {author} {\bibfnamefont {R.}~\bibnamefont {Brett}}, \bibinfo
  {author} {\bibfnamefont {A.}~\bibnamefont {Alexandru}},\ and\ \bibinfo
  {author} {\bibfnamefont {M.}~\bibnamefont {D{\"o}ring}},\ }\href@noop {} {\
  (\bibinfo {year} {2019})},\ \Eprint {https://arxiv.org/abs/1911.09047}
  {arXiv:1911.09047 [hep-lat]}\BibitemShut {NoStop}%
%%CITATION = ARXIV:1911.09047;%%
\bibitem [{\citenamefont {Khuri}\ and\ \citenamefont
  {Treiman}(1960)}]{Khuri:1960zz}%
  \BibitemOpen
  \bibfield  {author} {\bibinfo {author} {\bibfnamefont {N.~N.}\ \bibnamefont
  {Khuri}}\ and\ \bibinfo {author} {\bibfnamefont {S.~B.}\ \bibnamefont
  {Treiman}},\ }\href {https://doi.org/10.1103/PhysRev.119.1115} {\bibfield
  {journal} {\bibinfo  {journal} {Phys. Rev.}\ }\textbf {\bibinfo {volume}
  {119}},\ \bibinfo {pages} {1115} (\bibinfo {year} {1960})}\BibitemShut
  {NoStop}%
%%CITATION = PHRVA,119,1115;%%
\bibitem [{\citenamefont {Guo}\ \emph {et~al.}(2015{\natexlab{a}})\citenamefont
  {Guo}, \citenamefont {Danilkin},\ and\ \citenamefont
  {Szczepaniak}}]{Guo:2014vya}%
  \BibitemOpen
  \bibfield  {author} {\bibinfo {author} {\bibfnamefont {P.}~\bibnamefont
  {Guo}}, \bibinfo {author} {\bibfnamefont {I.}~\bibnamefont {Danilkin}},\ and\
  \bibinfo {author} {\bibfnamefont {A.~P.}\ \bibnamefont {Szczepaniak}},\
  }\href {https://doi.org/10.1140/epja/i2015-15135-7} {\bibfield  {journal}
  {\bibinfo  {journal} {Eur. Phys. J. A}\ }\textbf {\bibinfo {volume} {51}},\
  \bibinfo {pages} {135} (\bibinfo {year} {2015}{\natexlab{a}})},\ \Eprint
  {https://arxiv.org/abs/1409.8652} {arXiv:1409.8652 [hep-ph]}\BibitemShut
  {NoStop}%
\bibitem [{\citenamefont {Guo}\ \emph {et~al.}(2015{\natexlab{b}})\citenamefont
  {Guo}, \citenamefont {Danilkin}, \citenamefont {Schott}, \citenamefont
  {Fern\'{a}ndez-Ram\'{i}rez}, \citenamefont {Mathieu},\ and\ \citenamefont
  {Szczepaniak}}]{Guo:2015zqa}%
  \BibitemOpen
  \bibfield  {author} {\bibinfo {author} {\bibfnamefont {P.}~\bibnamefont
  {Guo}}, \bibinfo {author} {\bibfnamefont {I.~V.}\ \bibnamefont {Danilkin}},
  \bibinfo {author} {\bibfnamefont {D.}~\bibnamefont {Schott}}, \bibinfo
  {author} {\bibfnamefont {C.}~\bibnamefont {Fern\'{a}ndez-Ram\'{i}rez}},
  \bibinfo {author} {\bibfnamefont {V.}~\bibnamefont {Mathieu}},\ and\ \bibinfo
  {author} {\bibfnamefont {A.~P.}\ \bibnamefont {Szczepaniak}},\ }\href
  {https://doi.org/10.1103/PhysRevD.92.054016} {\bibfield  {journal} {\bibinfo
  {journal} {Phys. Rev.}\ }\textbf {\bibinfo {volume} {D92}},\ \bibinfo {pages}
  {054016} (\bibinfo {year} {2015}{\natexlab{b}})},\ \Eprint
  {https://arxiv.org/abs/1505.01715} {arXiv:1505.01715 [hep-ph]}\BibitemShut
  {NoStop}%
%%CITATION = ARXIV:1505.01715;%%
\bibitem [{\citenamefont {Guo}\ \emph {et~al.}(2017)\citenamefont {Guo},
  \citenamefont {Danilkin}, \citenamefont {Fern\'{a}ndez-Ram\'{i}rez},
  \citenamefont {Mathieu},\ and\ \citenamefont {Szczepaniak}}]{Guo:2016wsi}%
  \BibitemOpen
  \bibfield  {author} {\bibinfo {author} {\bibfnamefont {P.}~\bibnamefont
  {Guo}}, \bibinfo {author} {\bibfnamefont {I.~V.}\ \bibnamefont {Danilkin}},
  \bibinfo {author} {\bibfnamefont {C.}~\bibnamefont
  {Fern\'{a}ndez-Ram\'{i}rez}}, \bibinfo {author} {\bibfnamefont
  {V.}~\bibnamefont {Mathieu}},\ and\ \bibinfo {author} {\bibfnamefont {A.~P.}\
  \bibnamefont {Szczepaniak}},\ }\href
  {https://doi.org/10.1016/j.physletb.2017.05.092} {\bibfield  {journal}
  {\bibinfo  {journal} {Phys. Lett.}\ }\textbf {\bibinfo {volume} {B771}},\
  \bibinfo {pages} {497} (\bibinfo {year} {2017})},\ \Eprint
  {https://arxiv.org/abs/1608.01447} {arXiv:1608.01447 [hep-ph]}\BibitemShut
  {NoStop}%
%%CITATION = ARXIV:1608.01447;%%
\bibitem [{\citenamefont {Colangelo}\ \emph {et~al.}(2017)\citenamefont
  {Colangelo}, \citenamefont {Lanz}, \citenamefont {Leutwyler},\ and\
  \citenamefont {Passemar}}]{Colangelo:2016jmc}%
  \BibitemOpen
  \bibfield  {author} {\bibinfo {author} {\bibfnamefont {G.}~\bibnamefont
  {Colangelo}}, \bibinfo {author} {\bibfnamefont {S.}~\bibnamefont {Lanz}},
  \bibinfo {author} {\bibfnamefont {H.}~\bibnamefont {Leutwyler}},\ and\
  \bibinfo {author} {\bibfnamefont {E.}~\bibnamefont {Passemar}},\ }\href
  {https://doi.org/10.1103/PhysRevLett.118.022001} {\bibfield  {journal}
  {\bibinfo  {journal} {Phys. Rev. Lett.}\ }\textbf {\bibinfo {volume} {118}},\
  \bibinfo {pages} {022001} (\bibinfo {year} {2017})},\ \Eprint
  {https://arxiv.org/abs/1610.03494} {arXiv:1610.03494 [hep-ph]}\BibitemShut
  {NoStop}%
%%CITATION = ARXIV:1610.03494;%%
\bibitem [{\citenamefont {Colangelo}\ \emph {et~al.}(2018)\citenamefont
  {Colangelo}, \citenamefont {Lanz}, \citenamefont {Leutwyler},\ and\
  \citenamefont {Passemar}}]{Colangelo:2018jxw}%
  \BibitemOpen
  \bibfield  {author} {\bibinfo {author} {\bibfnamefont {G.}~\bibnamefont
  {Colangelo}}, \bibinfo {author} {\bibfnamefont {S.}~\bibnamefont {Lanz}},
  \bibinfo {author} {\bibfnamefont {H.}~\bibnamefont {Leutwyler}},\ and\
  \bibinfo {author} {\bibfnamefont {E.}~\bibnamefont {Passemar}},\ }\href
  {https://doi.org/10.1140/epjc/s10052-018-6377-9} {\bibfield  {journal}
  {\bibinfo  {journal} {Eur. Phys. J.}\ }\textbf {\bibinfo {volume} {C78}},\
  \bibinfo {pages} {947} (\bibinfo {year} {2018})},\ \Eprint
  {https://arxiv.org/abs/1807.11937} {arXiv:1807.11937 [hep-ph]}\BibitemShut
  {NoStop}%
%%CITATION = ARXIV:1807.11937;%%
\bibitem [{\citenamefont {Albaladejo}\ and\ \citenamefont
  {Moussallam}(2017)}]{Albaladejo:2017hhj}%
  \BibitemOpen
  \bibfield  {author} {\bibinfo {author} {\bibfnamefont {M.}~\bibnamefont
  {Albaladejo}}\ and\ \bibinfo {author} {\bibfnamefont {B.}~\bibnamefont
  {Moussallam}},\ }\href {https://doi.org/10.1140/epjc/s10052-017-5052-x}
  {\bibfield  {journal} {\bibinfo  {journal} {Eur. Phys. J.}\ }\textbf
  {\bibinfo {volume} {C77}},\ \bibinfo {pages} {508} (\bibinfo {year}
  {2017})},\ \Eprint {https://arxiv.org/abs/1702.04931} {arXiv:1702.04931
  [hep-ph]}\BibitemShut {NoStop}%
%%CITATION = ARXIV:1702.04931;%%
\bibitem [{\citenamefont {Gasser}\ and\ \citenamefont
  {Rusetsky}(2018)}]{Gasser:2018qtg}%
  \BibitemOpen
  \bibfield  {author} {\bibinfo {author} {\bibfnamefont {J.}~\bibnamefont
  {Gasser}}\ and\ \bibinfo {author} {\bibfnamefont {A.}~\bibnamefont
  {Rusetsky}},\ }\href {https://doi.org/10.1140/epjc/s10052-018-6378-8}
  {\bibfield  {journal} {\bibinfo  {journal} {Eur. Phys. J.}\ }\textbf
  {\bibinfo {volume} {C78}},\ \bibinfo {pages} {906} (\bibinfo {year}
  {2018})},\ \Eprint {https://arxiv.org/abs/1809.06399} {arXiv:1809.06399
  [hep-ph]}\BibitemShut {NoStop}%
%%CITATION = ARXIV:1809.06399;%%
\bibitem [{\citenamefont {Niecknig}\ \emph {et~al.}(2012)\citenamefont
  {Niecknig}, \citenamefont {Kubis},\ and\ \citenamefont
  {Schneider}}]{Niecknig:2012sj}%
  \BibitemOpen
  \bibfield  {author} {\bibinfo {author} {\bibfnamefont {F.}~\bibnamefont
  {Niecknig}}, \bibinfo {author} {\bibfnamefont {B.}~\bibnamefont {Kubis}},\
  and\ \bibinfo {author} {\bibfnamefont {S.~P.}\ \bibnamefont {Schneider}},\
  }\href {https://doi.org/10.1140/epjc/s10052-012-2014-1} {\bibfield  {journal}
  {\bibinfo  {journal} {Eur. Phys. J.}\ }\textbf {\bibinfo {volume} {C72}},\
  \bibinfo {pages} {2014} (\bibinfo {year} {2012})},\ \Eprint
  {https://arxiv.org/abs/1203.2501} {arXiv:1203.2501 [hep-ph]}\BibitemShut
  {NoStop}%
%%CITATION = ARXIV:1203.2501;%%
\bibitem [{\citenamefont {Danilkin}\ \emph {et~al.}(2015)\citenamefont
  {Danilkin}, \citenamefont {Fern\'andez-Ram\'irez}, \citenamefont {Guo},
  \citenamefont {Mathieu}, \citenamefont {Schott}, \citenamefont {Shi},\ and\
  \citenamefont {Szczepaniak}}]{Danilkin:2014cra}%
  \BibitemOpen
  \bibfield  {author} {\bibinfo {author} {\bibfnamefont {I.~V.}\ \bibnamefont
  {Danilkin}}, \bibinfo {author} {\bibfnamefont {C.}~\bibnamefont
  {Fern\'andez-Ram\'irez}}, \bibinfo {author} {\bibfnamefont {P.}~\bibnamefont
  {Guo}}, \bibinfo {author} {\bibfnamefont {V.}~\bibnamefont {Mathieu}},
  \bibinfo {author} {\bibfnamefont {D.}~\bibnamefont {Schott}}, \bibinfo
  {author} {\bibfnamefont {M.}~\bibnamefont {Shi}},\ and\ \bibinfo {author}
  {\bibfnamefont {A.~P.}\ \bibnamefont {Szczepaniak}},\ }\href
  {https://doi.org/10.1103/PhysRevD.91.094029} {\bibfield  {journal} {\bibinfo
  {journal} {Phys. Rev.}\ }\textbf {\bibinfo {volume} {D91}},\ \bibinfo {pages}
  {094029} (\bibinfo {year} {2015})},\ \Eprint
  {https://arxiv.org/abs/1409.7708} {arXiv:1409.7708 [hep-ph]}\BibitemShut
  {NoStop}%
%%CITATION = ARXIV:1409.7708;%%
\bibitem [{\citenamefont {Niecknig}\ and\ \citenamefont
  {Kubis}(2015)}]{Niecknig:2015ija}%
  \BibitemOpen
  \bibfield  {author} {\bibinfo {author} {\bibfnamefont {F.}~\bibnamefont
  {Niecknig}}\ and\ \bibinfo {author} {\bibfnamefont {B.}~\bibnamefont
  {Kubis}},\ }\href {https://doi.org/10.1007/JHEP10(2015)142} {\bibfield
  {journal} {\bibinfo  {journal} {JHEP}\ }\textbf {\bibinfo {volume} {10}},\
  \bibinfo {pages} {142}},\ \Eprint {https://arxiv.org/abs/1509.03188}
  {arXiv:1509.03188 [hep-ph]}\BibitemShut {NoStop}%
%%CITATION = ARXIV:1509.03188;%%
\bibitem [{\citenamefont {Isken}\ \emph {et~al.}(2017)\citenamefont {Isken},
  \citenamefont {Kubis}, \citenamefont {Schneider},\ and\ \citenamefont
  {Stoffer}}]{Isken:2017dkw}%
  \BibitemOpen
  \bibfield  {author} {\bibinfo {author} {\bibfnamefont {T.}~\bibnamefont
  {Isken}}, \bibinfo {author} {\bibfnamefont {B.}~\bibnamefont {Kubis}},
  \bibinfo {author} {\bibfnamefont {S.~P.}\ \bibnamefont {Schneider}},\ and\
  \bibinfo {author} {\bibfnamefont {P.}~\bibnamefont {Stoffer}},\ }\href
  {https://doi.org/10.1140/epjc/s10052-017-5024-1} {\bibfield  {journal}
  {\bibinfo  {journal} {Eur. Phys. J.}\ }\textbf {\bibinfo {volume} {C77}},\
  \bibinfo {pages} {489} (\bibinfo {year} {2017})},\ \Eprint
  {https://arxiv.org/abs/1705.04339} {arXiv:1705.04339 [hep-ph]}\BibitemShut
  {NoStop}%
%%CITATION = ARXIV:1705.04339;%%
\bibitem [{\citenamefont {Niecknig}\ and\ \citenamefont
  {Kubis}(2018)}]{Niecknig:2017ylb}%
  \BibitemOpen
  \bibfield  {author} {\bibinfo {author} {\bibfnamefont {F.}~\bibnamefont
  {Niecknig}}\ and\ \bibinfo {author} {\bibfnamefont {B.}~\bibnamefont
  {Kubis}},\ }\href {https://doi.org/10.1016/j.physletb.2018.03.048} {\bibfield
   {journal} {\bibinfo  {journal} {Phys. Lett.}\ }\textbf {\bibinfo {volume}
  {B780}},\ \bibinfo {pages} {471} (\bibinfo {year} {2018})},\ \Eprint
  {https://arxiv.org/abs/1708.00446} {arXiv:1708.00446 [hep-ph]}\BibitemShut
  {NoStop}%
%%CITATION = ARXIV:1708.00446;%%
\bibitem [{\citenamefont {Albaladejo}\ \emph {et~al.}(2020)\citenamefont
  {Albaladejo}, \citenamefont {Winney}, \citenamefont {Danilkin}, \citenamefont
  {Fern\'andez-Ram\'irez}, \citenamefont {Mathieu}, \citenamefont {Mikhasenko},
  \citenamefont {Pilloni}, \citenamefont {Silva-Castro},\ and\ \citenamefont
  {Szczepaniak}}]{Albaladejo:2019huw}%
  \BibitemOpen
  \bibfield  {author} {\bibinfo {author} {\bibfnamefont {M.}~\bibnamefont
  {Albaladejo}}, \bibinfo {author} {\bibfnamefont {D.}~\bibnamefont {Winney}},
  \bibinfo {author} {\bibfnamefont {I.}~\bibnamefont {Danilkin}}, \bibinfo
  {author} {\bibfnamefont {C.}~\bibnamefont {Fern\'andez-Ram\'irez}}, \bibinfo
  {author} {\bibfnamefont {V.}~\bibnamefont {Mathieu}}, \bibinfo {author}
  {\bibfnamefont {M.}~\bibnamefont {Mikhasenko}}, \bibinfo {author}
  {\bibfnamefont {A.}~\bibnamefont {Pilloni}}, \bibinfo {author} {\bibfnamefont
  {J.}~\bibnamefont {Silva-Castro}},\ and\ \bibinfo {author} {\bibfnamefont
  {A.}~\bibnamefont {Szczepaniak}} (\bibinfo {collaboration} {JPAC}),\ }\href
  {https://doi.org/10.1103/PhysRevD.101.054018} {\bibfield  {journal} {\bibinfo
   {journal} {Phys. Rev. D}\ }\textbf {\bibinfo {volume} {101}},\ \bibinfo
  {pages} {054018} (\bibinfo {year} {2020})},\ \Eprint
  {https://arxiv.org/abs/1910.03107} {arXiv:1910.03107 [hep-ph]}\BibitemShut
  {NoStop}%
\bibitem [{\citenamefont {Mikhasenko}\ \emph {et~al.}(2020)\citenamefont
  {Mikhasenko} \emph {et~al.}}]{Mikhasenko:2019rjf}%
  \BibitemOpen
  \bibfield  {author} {\bibinfo {author} {\bibfnamefont {M.}~\bibnamefont
  {Mikhasenko}} \emph {et~al.} (\bibinfo {collaboration} {JPAC}),\ }\href
  {https://doi.org/10.1103/PhysRevD.101.034033} {\bibfield  {journal} {\bibinfo
   {journal} {Phys. Rev.}\ }\textbf {\bibinfo {volume} {D101}},\ \bibinfo
  {pages} {034033} (\bibinfo {year} {2020})},\ \Eprint
  {https://arxiv.org/abs/1910.04566} {arXiv:1910.04566 [hep-ph]}\BibitemShut
  {NoStop}%
%%CITATION = ARXIV:1910.04566;%%
\bibitem [{\citenamefont {Garcia-Martin}\ \emph {et~al.}(2011)\citenamefont
  {Garcia-Martin}, \citenamefont {Kaminski}, \citenamefont {Pelaez},
  \citenamefont {Ruiz~de Elvira},\ and\ \citenamefont
  {Yndurain}}]{GarciaMartin:2011cn}%
  \BibitemOpen
  \bibfield  {author} {\bibinfo {author} {\bibfnamefont {R.}~\bibnamefont
  {Garcia-Martin}}, \bibinfo {author} {\bibfnamefont {R.}~\bibnamefont
  {Kaminski}}, \bibinfo {author} {\bibfnamefont {J.~R.}\ \bibnamefont
  {Pelaez}}, \bibinfo {author} {\bibfnamefont {J.}~\bibnamefont {Ruiz~de
  Elvira}},\ and\ \bibinfo {author} {\bibfnamefont {F.~J.}\ \bibnamefont
  {Yndurain}},\ }\href {https://doi.org/10.1103/PhysRevD.83.074004} {\bibfield
  {journal} {\bibinfo  {journal} {Phys. Rev.}\ }\textbf {\bibinfo {volume}
  {D83}},\ \bibinfo {pages} {074004} (\bibinfo {year} {2011})},\ \Eprint
  {https://arxiv.org/abs/1102.2183} {arXiv:1102.2183 [hep-ph]}\BibitemShut
  {NoStop}%
%%CITATION = ARXIV:1102.2183;%%
\bibitem [{\citenamefont {Dax}\ \emph {et~al.}(2018)\citenamefont {Dax},
  \citenamefont {Isken},\ and\ \citenamefont {Kubis}}]{Dax:2018rvs}%
  \BibitemOpen
  \bibfield  {author} {\bibinfo {author} {\bibfnamefont {M.}~\bibnamefont
  {Dax}}, \bibinfo {author} {\bibfnamefont {T.}~\bibnamefont {Isken}},\ and\
  \bibinfo {author} {\bibfnamefont {B.}~\bibnamefont {Kubis}},\ }\href
  {https://doi.org/10.1140/epjc/s10052-018-6346-3} {\bibfield  {journal}
  {\bibinfo  {journal} {Eur. Phys. J. C}\ }\textbf {\bibinfo {volume} {78}},\
  \bibinfo {pages} {859} (\bibinfo {year} {2018})},\ \Eprint
  {https://arxiv.org/abs/1808.08957} {arXiv:1808.08957 [hep-ph]}\BibitemShut
  {NoStop}%
\bibitem [{\citenamefont {Aloisio}\ \emph {et~al.}(2003)\citenamefont {Aloisio}
  \emph {et~al.}}]{Aloisio:2003ur}%
  \BibitemOpen
  \bibfield  {author} {\bibinfo {author} {\bibfnamefont {A.}~\bibnamefont
  {Aloisio}} \emph {et~al.} (\bibinfo {collaboration} {KLOE}),\ }\href
  {https://doi.org/10.1016/j.physletb.2005.01.092} {\bibfield  {journal}
  {\bibinfo  {journal} {Phys. Lett. B}\ }\textbf {\bibinfo {volume} {561}},\
  \bibinfo {pages} {55} (\bibinfo {year} {2003})},\ \bibinfo {note} {[Erratum:
  Phys.Lett.B 609, 449--450 (2005)]},\ \Eprint
  {https://arxiv.org/abs/hep-ex/0303016} {arXiv:hep-ex/0303016}\BibitemShut
  {NoStop}%
\bibitem [{\citenamefont {Akhmetshin}\ \emph {et~al.}(2006)\citenamefont
  {Akhmetshin} \emph {et~al.}}]{Akhmetshin:2006sc}%
  \BibitemOpen
  \bibfield  {author} {\bibinfo {author} {\bibfnamefont {R.}~\bibnamefont
  {Akhmetshin}} \emph {et~al.},\ }\href
  {https://doi.org/10.1016/j.physletb.2006.09.041} {\bibfield  {journal}
  {\bibinfo  {journal} {Phys. Lett. B}\ }\textbf {\bibinfo {volume} {642}},\
  \bibinfo {pages} {203} (\bibinfo {year} {2006})}\BibitemShut {NoStop}%
\bibitem [{\citenamefont {Adlarson}\ \emph
  {et~al.}(2017{\natexlab{a}})\citenamefont {Adlarson} \emph
  {et~al.}}]{Adlarson:2016wkw}%
  \BibitemOpen
  \bibfield  {author} {\bibinfo {author} {\bibfnamefont {P.}~\bibnamefont
  {Adlarson}} \emph {et~al.} (\bibinfo {collaboration} {WASA-at-COSY}),\ }\href
  {https://doi.org/10.1016/j.physletb.2017.03.050} {\bibfield  {journal}
  {\bibinfo  {journal} {Phys. Lett. B}\ }\textbf {\bibinfo {volume} {770}},\
  \bibinfo {pages} {418} (\bibinfo {year} {2017}{\natexlab{a}})},\ \Eprint
  {https://arxiv.org/abs/1610.02187} {arXiv:1610.02187 [nucl-ex]}\BibitemShut
  {NoStop}%
\bibitem [{\citenamefont {Ablikim}\ \emph {et~al.}(2018)\citenamefont {Ablikim}
  \emph {et~al.}}]{Ablikim:2018yen}%
  \BibitemOpen
  \bibfield  {author} {\bibinfo {author} {\bibfnamefont {M.}~\bibnamefont
  {Ablikim}} \emph {et~al.} (\bibinfo {collaboration} {BESIII}),\ }\href
  {https://doi.org/10.1103/PhysRevD.98.112007} {\bibfield  {journal} {\bibinfo
  {journal} {Phys. Rev. D}\ }\textbf {\bibinfo {volume} {98}},\ \bibinfo
  {pages} {112007} (\bibinfo {year} {2018})},\ \Eprint
  {https://arxiv.org/abs/1811.03817} {arXiv:1811.03817 [hep-ex]}\BibitemShut
  {NoStop}%
\bibitem [{\citenamefont {Adlarson}\ \emph
  {et~al.}(2017{\natexlab{b}})\citenamefont {Adlarson} \emph
  {et~al.}}]{Adlarson:2016hpp}%
  \BibitemOpen
  \bibfield  {author} {\bibinfo {author} {\bibfnamefont {P.}~\bibnamefont
  {Adlarson}} \emph {et~al.},\ }\href
  {https://doi.org/10.1103/PhysRevC.95.035208} {\bibfield  {journal} {\bibinfo
  {journal} {Phys. Rev. C}\ }\textbf {\bibinfo {volume} {95}},\ \bibinfo
  {pages} {035208} (\bibinfo {year} {2017}{\natexlab{b}})},\ \Eprint
  {https://arxiv.org/abs/1609.04503} {arXiv:1609.04503 [hep-ex]}\BibitemShut
  {NoStop}%
\bibitem [{\citenamefont {Arnaldi}\ \emph {et~al.}(2009)\citenamefont {Arnaldi}
  \emph {et~al.}}]{Arnaldi:2009aa}%
  \BibitemOpen
  \bibfield  {author} {\bibinfo {author} {\bibfnamefont {R.}~\bibnamefont
  {Arnaldi}} \emph {et~al.} (\bibinfo {collaboration} {NA60}),\ }\href
  {https://doi.org/10.1016/j.physletb.2009.05.029} {\bibfield  {journal}
  {\bibinfo  {journal} {Phys. Lett. B}\ }\textbf {\bibinfo {volume} {677}},\
  \bibinfo {pages} {260} (\bibinfo {year} {2009})},\ \Eprint
  {https://arxiv.org/abs/0902.2547} {arXiv:0902.2547 [hep-ph]}\BibitemShut
  {NoStop}%
\bibitem [{\citenamefont {Arnaldi}\ \emph {et~al.}(2016)\citenamefont {Arnaldi}
  \emph {et~al.}}]{Arnaldi:2016pzu}%
  \BibitemOpen
  \bibfield  {author} {\bibinfo {author} {\bibfnamefont {R.}~\bibnamefont
  {Arnaldi}} \emph {et~al.} (\bibinfo {collaboration} {NA60}),\ }\href
  {https://doi.org/10.1016/j.physletb.2016.04.013} {\bibfield  {journal}
  {\bibinfo  {journal} {Phys. Lett. B}\ }\textbf {\bibinfo {volume} {757}},\
  \bibinfo {pages} {437} (\bibinfo {year} {2016})},\ \Eprint
  {https://arxiv.org/abs/1608.07898} {arXiv:1608.07898 [hep-ex]}\BibitemShut
  {NoStop}%
\bibitem [{\citenamefont {Jegerlehner}(2017)}]{Jegerlehner:2017gek}%
  \BibitemOpen
  \bibfield  {author} {\bibinfo {author} {\bibfnamefont {F.}~\bibnamefont
  {Jegerlehner}},\ }\href {https://doi.org/10.1007/978-3-319-63577-4} {\emph
  {\bibinfo {title} {{The Anomalous Magnetic Moment of the Muon}}}},\ Vol.\
  \bibinfo {volume} {274}\ (\bibinfo  {publisher} {Springer},\ \bibinfo
  {address} {Cham},\ \bibinfo {year} {2017})\BibitemShut {NoStop}%
\bibitem [{\citenamefont {Keshavarzi}\ \emph {et~al.}(2018)\citenamefont
  {Keshavarzi}, \citenamefont {Nomura},\ and\ \citenamefont
  {Teubner}}]{Keshavarzi:2018mgv}%
  \BibitemOpen
  \bibfield  {author} {\bibinfo {author} {\bibfnamefont {A.}~\bibnamefont
  {Keshavarzi}}, \bibinfo {author} {\bibfnamefont {D.}~\bibnamefont {Nomura}},\
  and\ \bibinfo {author} {\bibfnamefont {T.}~\bibnamefont {Teubner}},\ }\href
  {https://doi.org/10.1103/PhysRevD.97.114025} {\bibfield  {journal} {\bibinfo
  {journal} {Phys. Rev. D}\ }\textbf {\bibinfo {volume} {97}},\ \bibinfo
  {pages} {114025} (\bibinfo {year} {2018})},\ \Eprint
  {https://arxiv.org/abs/1802.02995} {arXiv:1802.02995 [hep-ph]}\BibitemShut
  {NoStop}%
\bibitem [{\citenamefont {Davier}\ \emph {et~al.}(2017)\citenamefont {Davier},
  \citenamefont {Hoecker}, \citenamefont {Malaescu},\ and\ \citenamefont
  {Zhang}}]{Davier:2017zfy}%
  \BibitemOpen
  \bibfield  {author} {\bibinfo {author} {\bibfnamefont {M.}~\bibnamefont
  {Davier}}, \bibinfo {author} {\bibfnamefont {A.}~\bibnamefont {Hoecker}},
  \bibinfo {author} {\bibfnamefont {B.}~\bibnamefont {Malaescu}},\ and\
  \bibinfo {author} {\bibfnamefont {Z.}~\bibnamefont {Zhang}},\ }\href
  {https://doi.org/10.1140/epjc/s10052-017-5161-6} {\bibfield  {journal}
  {\bibinfo  {journal} {Eur. Phys. J. C}\ }\textbf {\bibinfo {volume} {77}},\
  \bibinfo {pages} {827} (\bibinfo {year} {2017})},\ \Eprint
  {https://arxiv.org/abs/1706.09436} {arXiv:1706.09436 [hep-ph]}\BibitemShut
  {NoStop}%
\bibitem [{\citenamefont {Danilkin}\ \emph {et~al.}(2019)\citenamefont
  {Danilkin}, \citenamefont {Redmer},\ and\ \citenamefont
  {Vanderhaeghen}}]{Danilkin:2019mhd}%
  \BibitemOpen
  \bibfield  {author} {\bibinfo {author} {\bibfnamefont {I.}~\bibnamefont
  {Danilkin}}, \bibinfo {author} {\bibfnamefont {C.~F.}\ \bibnamefont
  {Redmer}},\ and\ \bibinfo {author} {\bibfnamefont {M.}~\bibnamefont
  {Vanderhaeghen}},\ }\href {https://doi.org/10.1016/j.ppnp.2019.05.002}
  {\bibfield  {journal} {\bibinfo  {journal} {Prog. Part. Nucl. Phys.}\
  }\textbf {\bibinfo {volume} {107}},\ \bibinfo {pages} {20} (\bibinfo {year}
  {2019})},\ \Eprint {https://arxiv.org/abs/1901.10346} {arXiv:1901.10346
  [hep-ph]}\BibitemShut {NoStop}%
\bibitem [{\citenamefont {Tanabashi}\ \emph {et~al.}(2018)\citenamefont
  {Tanabashi} \emph {et~al.}}]{Tanabashi:2018oca}%
  \BibitemOpen
  \bibfield  {author} {\bibinfo {author} {\bibfnamefont {M.}~\bibnamefont
  {Tanabashi}} \emph {et~al.} (\bibinfo {collaboration} {Particle Data
  Group}),\ }\href {https://doi.org/10.1103/PhysRevD.98.030001} {\bibfield
  {journal} {\bibinfo  {journal} {Phys. Rev.}\ }\textbf {\bibinfo {volume}
  {D98}},\ \bibinfo {pages} {030001} (\bibinfo {year} {2018})}\BibitemShut
  {NoStop}%
%%CITATION = PHRVA,D98,030001;%%
\bibitem [{\citenamefont {Lee~Roberts}(2011)}]{LeeRoberts:2011zz}%
  \BibitemOpen
  \bibfield  {author} {\bibinfo {author} {\bibfnamefont {B.}~\bibnamefont
  {Lee~Roberts}} (\bibinfo {collaboration} {Fermilab P989}),\ }\href
  {https://doi.org/10.1016/j.nuclphysbps.2011.06.038} {\bibfield  {journal}
  {\bibinfo  {journal} {Nucl. Phys. B Proc. Suppl.}\ }\textbf {\bibinfo
  {volume} {218}},\ \bibinfo {pages} {237} (\bibinfo {year}
  {2011})}\BibitemShut {NoStop}%
\bibitem [{\citenamefont {Grange}\ \emph {et~al.}(2015)\citenamefont {Grange}
  \emph {et~al.}}]{Grange:2015fou}%
  \BibitemOpen
  \bibfield  {author} {\bibinfo {author} {\bibfnamefont {J.}~\bibnamefont
  {Grange}} \emph {et~al.} (\bibinfo {collaboration} {Muon g-2}),\ }\href@noop
  {} {\  (\bibinfo {year} {2015})},\ \Eprint {https://arxiv.org/abs/1501.06858}
  {arXiv:1501.06858 [physics.ins-det]}\BibitemShut {NoStop}%
\bibitem [{\citenamefont {Iinuma}(2011)}]{Iinuma:2011zz}%
  \BibitemOpen
  \bibfield  {author} {\bibinfo {author} {\bibfnamefont {H.}~\bibnamefont
  {Iinuma}} (\bibinfo {collaboration} {J-PARC muon g-2/EDM}),\ }\href
  {https://doi.org/10.1088/1742-6596/295/1/012032} {\bibfield  {journal}
  {\bibinfo  {journal} {J. Phys. Conf. Ser.}\ }\textbf {\bibinfo {volume}
  {295}},\ \bibinfo {pages} {012032} (\bibinfo {year} {2011})}\BibitemShut
  {NoStop}%
\bibitem [{\citenamefont {Hoferichter}\ \emph {et~al.}(2014)\citenamefont
  {Hoferichter}, \citenamefont {Kubis}, \citenamefont {Leupold}, \citenamefont
  {Niecknig},\ and\ \citenamefont {Schneider}}]{Hoferichter:2014vra}%
  \BibitemOpen
  \bibfield  {author} {\bibinfo {author} {\bibfnamefont {M.}~\bibnamefont
  {Hoferichter}}, \bibinfo {author} {\bibfnamefont {B.}~\bibnamefont {Kubis}},
  \bibinfo {author} {\bibfnamefont {S.}~\bibnamefont {Leupold}}, \bibinfo
  {author} {\bibfnamefont {F.}~\bibnamefont {Niecknig}},\ and\ \bibinfo
  {author} {\bibfnamefont {S.~P.}\ \bibnamefont {Schneider}},\ }\href
  {https://doi.org/10.1140/epjc/s10052-014-3180-0} {\bibfield  {journal}
  {\bibinfo  {journal} {Eur. Phys. J. C}\ }\textbf {\bibinfo {volume} {74}},\
  \bibinfo {pages} {3180} (\bibinfo {year} {2014})},\ \Eprint
  {https://arxiv.org/abs/1410.4691} {arXiv:1410.4691 [hep-ph]}\BibitemShut
  {NoStop}%
\bibitem [{\citenamefont {Hoferichter}\ \emph
  {et~al.}(2018{\natexlab{a}})\citenamefont {Hoferichter}, \citenamefont
  {Hoid}, \citenamefont {Kubis}, \citenamefont {Leupold},\ and\ \citenamefont
  {Schneider}}]{Hoferichter:2018kwz}%
  \BibitemOpen
  \bibfield  {author} {\bibinfo {author} {\bibfnamefont {M.}~\bibnamefont
  {Hoferichter}}, \bibinfo {author} {\bibfnamefont {B.-L.}\ \bibnamefont
  {Hoid}}, \bibinfo {author} {\bibfnamefont {B.}~\bibnamefont {Kubis}},
  \bibinfo {author} {\bibfnamefont {S.}~\bibnamefont {Leupold}},\ and\ \bibinfo
  {author} {\bibfnamefont {S.~P.}\ \bibnamefont {Schneider}},\ }\href
  {https://doi.org/10.1007/JHEP10(2018)141} {\bibfield  {journal} {\bibinfo
  {journal} {JHEP}\ }\textbf {\bibinfo {volume} {10}},\ \bibinfo {pages}
  {141}},\ \Eprint {https://arxiv.org/abs/1808.04823} {arXiv:1808.04823
  [hep-ph]}\BibitemShut {NoStop}%
\bibitem [{\citenamefont {Hoferichter}\ \emph
  {et~al.}(2018{\natexlab{b}})\citenamefont {Hoferichter}, \citenamefont
  {Hoid}, \citenamefont {Kubis}, \citenamefont {Leupold},\ and\ \citenamefont
  {Schneider}}]{Hoferichter:2018dmo}%
  \BibitemOpen
  \bibfield  {author} {\bibinfo {author} {\bibfnamefont {M.}~\bibnamefont
  {Hoferichter}}, \bibinfo {author} {\bibfnamefont {B.-L.}\ \bibnamefont
  {Hoid}}, \bibinfo {author} {\bibfnamefont {B.}~\bibnamefont {Kubis}},
  \bibinfo {author} {\bibfnamefont {S.}~\bibnamefont {Leupold}},\ and\ \bibinfo
  {author} {\bibfnamefont {S.~P.}\ \bibnamefont {Schneider}},\ }\href
  {https://doi.org/10.1103/PhysRevLett.121.112002} {\bibfield  {journal}
  {\bibinfo  {journal} {Phys. Rev. Lett.}\ }\textbf {\bibinfo {volume} {121}},\
  \bibinfo {pages} {112002} (\bibinfo {year} {2018}{\natexlab{b}})},\ \Eprint
  {https://arxiv.org/abs/1805.01471} {arXiv:1805.01471 [hep-ph]}\BibitemShut
  {NoStop}%
\bibitem [{\citenamefont {Hoferichter}\ \emph {et~al.}(2019)\citenamefont
  {Hoferichter}, \citenamefont {Hoid},\ and\ \citenamefont
  {Kubis}}]{Hoferichter:2019mqg}%
  \BibitemOpen
  \bibfield  {author} {\bibinfo {author} {\bibfnamefont {M.}~\bibnamefont
  {Hoferichter}}, \bibinfo {author} {\bibfnamefont {B.-L.}\ \bibnamefont
  {Hoid}},\ and\ \bibinfo {author} {\bibfnamefont {B.}~\bibnamefont {Kubis}},\
  }\href {https://doi.org/10.1007/JHEP08(2019)137} {\bibfield  {journal}
  {\bibinfo  {journal} {JHEP}\ }\textbf {\bibinfo {volume} {08}},\ \bibinfo
  {pages} {137}},\ \Eprint {https://arxiv.org/abs/1907.01556} {arXiv:1907.01556
  [hep-ph]}\BibitemShut {NoStop}%
\bibitem [{\citenamefont {Colangelo}\ \emph {et~al.}(2014)\citenamefont
  {Colangelo}, \citenamefont {Hoferichter}, \citenamefont {Kubis},
  \citenamefont {Procura},\ and\ \citenamefont {Stoffer}}]{Colangelo:2014pva}%
  \BibitemOpen
  \bibfield  {author} {\bibinfo {author} {\bibfnamefont {G.}~\bibnamefont
  {Colangelo}}, \bibinfo {author} {\bibfnamefont {M.}~\bibnamefont
  {Hoferichter}}, \bibinfo {author} {\bibfnamefont {B.}~\bibnamefont {Kubis}},
  \bibinfo {author} {\bibfnamefont {M.}~\bibnamefont {Procura}},\ and\ \bibinfo
  {author} {\bibfnamefont {P.}~\bibnamefont {Stoffer}},\ }\href
  {https://doi.org/10.1016/j.physletb.2014.09.021} {\bibfield  {journal}
  {\bibinfo  {journal} {Phys. Lett. B}\ }\textbf {\bibinfo {volume} {738}},\
  \bibinfo {pages} {6} (\bibinfo {year} {2014})},\ \Eprint
  {https://arxiv.org/abs/1408.2517} {arXiv:1408.2517 [hep-ph]}\BibitemShut
  {NoStop}%
\bibitem [{\citenamefont {Danilkin}\ and\ \citenamefont
  {Vanderhaeghen}(2019)}]{Danilkin:2018qfn}%
  \BibitemOpen
  \bibfield  {author} {\bibinfo {author} {\bibfnamefont {I.}~\bibnamefont
  {Danilkin}}\ and\ \bibinfo {author} {\bibfnamefont {M.}~\bibnamefont
  {Vanderhaeghen}},\ }\href {https://doi.org/10.1016/j.physletb.2018.12.047}
  {\bibfield  {journal} {\bibinfo  {journal} {Phys. Lett. B}\ }\textbf
  {\bibinfo {volume} {789}},\ \bibinfo {pages} {366} (\bibinfo {year}
  {2019})},\ \Eprint {https://arxiv.org/abs/1810.03669} {arXiv:1810.03669
  [hep-ph]}\BibitemShut {NoStop}%
\bibitem [{\citenamefont {Hoferichter}\ and\ \citenamefont
  {Stoffer}(2019)}]{Hoferichter:2019nlq}%
  \BibitemOpen
  \bibfield  {author} {\bibinfo {author} {\bibfnamefont {M.}~\bibnamefont
  {Hoferichter}}\ and\ \bibinfo {author} {\bibfnamefont {P.}~\bibnamefont
  {Stoffer}},\ }\href {https://doi.org/10.1007/JHEP07(2019)073} {\bibfield
  {journal} {\bibinfo  {journal} {JHEP}\ }\textbf {\bibinfo {volume} {07}},\
  \bibinfo {pages} {073}},\ \Eprint {https://arxiv.org/abs/1905.13198}
  {arXiv:1905.13198 [hep-ph]}\BibitemShut {NoStop}%
\bibitem [{\citenamefont {Danilkin}\ \emph {et~al.}(2020)\citenamefont
  {Danilkin}, \citenamefont {Deineka},\ and\ \citenamefont
  {Vanderhaeghen}}]{Danilkin:2019opj}%
  \BibitemOpen
  \bibfield  {author} {\bibinfo {author} {\bibfnamefont {I.}~\bibnamefont
  {Danilkin}}, \bibinfo {author} {\bibfnamefont {O.}~\bibnamefont {Deineka}},\
  and\ \bibinfo {author} {\bibfnamefont {M.}~\bibnamefont {Vanderhaeghen}},\
  }\href {https://doi.org/10.1103/PhysRevD.101.054008} {\bibfield  {journal}
  {\bibinfo  {journal} {Phys. Rev. D}\ }\textbf {\bibinfo {volume} {101}},\
  \bibinfo {pages} {054008} (\bibinfo {year} {2020})},\ \Eprint
  {https://arxiv.org/abs/1909.04158} {arXiv:1909.04158 [hep-ph]}\BibitemShut
  {NoStop}%
\bibitem [{\citenamefont {K{\"a}ll{\'e}n}(1964)}]{Kallen:1964lxa}%
  \BibitemOpen
  \bibfield  {author} {\bibinfo {author} {\bibfnamefont {G.}~\bibnamefont
  {K{\"a}ll{\'e}n}},\ }\href@noop {} {\emph {\bibinfo {title} {{Elementary
  particle physics}}}}\ (\bibinfo  {publisher} {Addison-Wesley},\ \bibinfo
  {address} {Reading, MA},\ \bibinfo {year} {1964})\BibitemShut {NoStop}%
%%CITATION = INSPIRE-1288389;%%
\bibitem [{\citenamefont {Kibble}(1960)}]{Kibble:1960zz}%
  \BibitemOpen
  \bibfield  {author} {\bibinfo {author} {\bibfnamefont {T.~W.~B.}\
  \bibnamefont {Kibble}},\ }\href {https://doi.org/10.1103/PhysRev.117.1159}
  {\bibfield  {journal} {\bibinfo  {journal} {Phys. Rev.}\ }\textbf {\bibinfo
  {volume} {117}},\ \bibinfo {pages} {1159} (\bibinfo {year}
  {1960})}\BibitemShut {NoStop}%
%%CITATION = PHRVA,117,1159;%%
\bibitem [{\citenamefont {Stern}\ \emph {et~al.}(1993)\citenamefont {Stern},
  \citenamefont {Sazdjian},\ and\ \citenamefont {Fuchs}}]{Stern:1993rg}%
  \BibitemOpen
  \bibfield  {author} {\bibinfo {author} {\bibfnamefont {J.}~\bibnamefont
  {Stern}}, \bibinfo {author} {\bibfnamefont {H.}~\bibnamefont {Sazdjian}},\
  and\ \bibinfo {author} {\bibfnamefont {N.~H.}\ \bibnamefont {Fuchs}},\ }\href
  {https://doi.org/10.1103/PhysRevD.47.3814} {\bibfield  {journal} {\bibinfo
  {journal} {Phys. Rev.}\ }\textbf {\bibinfo {volume} {D47}},\ \bibinfo {pages}
  {3814} (\bibinfo {year} {1993})},\ \Eprint
  {https://arxiv.org/abs/hep-ph/9301244} {arXiv:hep-ph/9301244
  [hep-ph]}\BibitemShut {NoStop}%
%%CITATION = HEP-PH/9301244;%%
\bibitem [{\citenamefont {Knecht}\ \emph {et~al.}(1995)\citenamefont {Knecht},
  \citenamefont {Moussallam}, \citenamefont {Stern},\ and\ \citenamefont
  {Fuchs}}]{Knecht:1995tr}%
  \BibitemOpen
  \bibfield  {author} {\bibinfo {author} {\bibfnamefont {M.}~\bibnamefont
  {Knecht}}, \bibinfo {author} {\bibfnamefont {B.}~\bibnamefont {Moussallam}},
  \bibinfo {author} {\bibfnamefont {J.}~\bibnamefont {Stern}},\ and\ \bibinfo
  {author} {\bibfnamefont {N.}~\bibnamefont {Fuchs}},\ }\href
  {https://doi.org/10.1016/0550-3213(95)00515-3} {\bibfield  {journal}
  {\bibinfo  {journal} {Nucl. Phys. B}\ }\textbf {\bibinfo {volume} {457}},\
  \bibinfo {pages} {513} (\bibinfo {year} {1995})},\ \Eprint
  {https://arxiv.org/abs/hep-ph/9507319} {arXiv:hep-ph/9507319}\BibitemShut
  {NoStop}%
\bibitem [{\citenamefont {Albaladejo}\ \emph {et~al.}(2018)\citenamefont
  {Albaladejo}, \citenamefont {Sherrill}, \citenamefont
  {Fern\'andez-Ram\'irez}, \citenamefont {Jackura}, \citenamefont {Mathieu},
  \citenamefont {Mikhasenko}, \citenamefont {Nys}, \citenamefont {Pilloni},\
  and\ \citenamefont {Szczepaniak}}]{Albaladejo:2018gif}%
  \BibitemOpen
  \bibfield  {author} {\bibinfo {author} {\bibfnamefont {M.}~\bibnamefont
  {Albaladejo}}, \bibinfo {author} {\bibfnamefont {N.}~\bibnamefont
  {Sherrill}}, \bibinfo {author} {\bibfnamefont {C.}~\bibnamefont
  {Fern\'andez-Ram\'irez}}, \bibinfo {author} {\bibfnamefont {A.}~\bibnamefont
  {Jackura}}, \bibinfo {author} {\bibfnamefont {V.}~\bibnamefont {Mathieu}},
  \bibinfo {author} {\bibfnamefont {M.}~\bibnamefont {Mikhasenko}}, \bibinfo
  {author} {\bibfnamefont {J.}~\bibnamefont {Nys}}, \bibinfo {author}
  {\bibfnamefont {A.}~\bibnamefont {Pilloni}},\ and\ \bibinfo {author}
  {\bibfnamefont {A.~P.}\ \bibnamefont {Szczepaniak}} (\bibinfo {collaboration}
  {JPAC}),\ }\href {https://doi.org/10.1140/epjc/s10052-018-6045-0} {\bibfield
  {journal} {\bibinfo  {journal} {Eur. Phys. J.}\ }\textbf {\bibinfo {volume}
  {C78}},\ \bibinfo {pages} {574} (\bibinfo {year} {2018})},\ \Eprint
  {https://arxiv.org/abs/1803.06027} {arXiv:1803.06027 [hep-ph]}\BibitemShut
  {NoStop}%
%%CITATION = ARXIV:1803.06027;%%
\bibitem [{\citenamefont {Bronzan}\ and\ \citenamefont
  {Kacser}(1963)}]{Bronzan:1963mby}%
  \BibitemOpen
  \bibfield  {author} {\bibinfo {author} {\bibfnamefont {J.~B.}\ \bibnamefont
  {Bronzan}}\ and\ \bibinfo {author} {\bibfnamefont {C.}~\bibnamefont
  {Kacser}},\ }\href {https://doi.org/10.1103/PhysRev.132.2703} {\bibfield
  {journal} {\bibinfo  {journal} {Phys.Rev.}\ }\textbf {\bibinfo {volume}
  {132}},\ \bibinfo {pages} {2703} (\bibinfo {year} {1963})}\BibitemShut
  {NoStop}%
%%CITATION = PHRVA,132,2703;%%
\bibitem [{\citenamefont {Omnes}(1958)}]{Omnes:1958hv}%
  \BibitemOpen
  \bibfield  {author} {\bibinfo {author} {\bibfnamefont {R.}~\bibnamefont
  {Omnes}},\ }\href {https://doi.org/10.1007/BF02747746} {\bibfield  {journal}
  {\bibinfo  {journal} {Nuovo Cim.}\ }\textbf {\bibinfo {volume} {8}},\
  \bibinfo {pages} {316} (\bibinfo {year} {1958})}\BibitemShut {NoStop}%
\bibitem [{\citenamefont {Gonz\`alez-Sol\'is}\ and\ \citenamefont
  {Roig}(2019)}]{Gonzalez-Solis:2019iod}%
  \BibitemOpen
  \bibfield  {author} {\bibinfo {author} {\bibfnamefont {S.}~\bibnamefont
  {Gonz\`alez-Sol\'is}}\ and\ \bibinfo {author} {\bibfnamefont
  {P.}~\bibnamefont {Roig}},\ }\href
  {https://doi.org/10.1140/epjc/s10052-019-6943-9} {\bibfield  {journal}
  {\bibinfo  {journal} {Eur. Phys. J. C}\ }\textbf {\bibinfo {volume} {79}},\
  \bibinfo {pages} {436} (\bibinfo {year} {2019})},\ \Eprint
  {https://arxiv.org/abs/1902.02273} {arXiv:1902.02273 [hep-ph]}\BibitemShut
  {NoStop}%
\bibitem [{\citenamefont {Froissart}(1961)}]{Froissart:1961ux}%
  \BibitemOpen
  \bibfield  {author} {\bibinfo {author} {\bibfnamefont {M.}~\bibnamefont
  {Froissart}},\ }\href {https://doi.org/10.1103/PhysRev.123.1053} {\bibfield
  {journal} {\bibinfo  {journal} {Phys. Rev.}\ }\textbf {\bibinfo {volume}
  {123}},\ \bibinfo {pages} {1053} (\bibinfo {year} {1961})}\BibitemShut
  {NoStop}%
\bibitem [{\citenamefont {Martin}(1963)}]{Martin:1962rt}%
  \BibitemOpen
  \bibfield  {author} {\bibinfo {author} {\bibfnamefont {A.}~\bibnamefont
  {Martin}},\ }\href {https://doi.org/10.1103/PhysRev.129.1432} {\bibfield
  {journal} {\bibinfo  {journal} {Phys. Rev.}\ }\textbf {\bibinfo {volume}
  {129}},\ \bibinfo {pages} {1432} (\bibinfo {year} {1963})}\BibitemShut
  {NoStop}%
\bibitem [{\citenamefont {Kambor}\ \emph {et~al.}(1996)\citenamefont {Kambor},
  \citenamefont {Wiesendanger},\ and\ \citenamefont {Wyler}}]{Kambor:1995yc}%
  \BibitemOpen
  \bibfield  {author} {\bibinfo {author} {\bibfnamefont {J.}~\bibnamefont
  {Kambor}}, \bibinfo {author} {\bibfnamefont {C.}~\bibnamefont
  {Wiesendanger}},\ and\ \bibinfo {author} {\bibfnamefont {D.}~\bibnamefont
  {Wyler}},\ }\href {https://doi.org/10.1016/0550-3213(95)00676-1} {\bibfield
  {journal} {\bibinfo  {journal} {Nucl. Phys.}\ }\textbf {\bibinfo {volume}
  {B465}},\ \bibinfo {pages} {215} (\bibinfo {year} {1996})},\ \Eprint
  {https://arxiv.org/abs/hep-ph/9509374} {arXiv:hep-ph/9509374
  [hep-ph]}\BibitemShut {NoStop}%
%%CITATION = HEP-PH/9509374;%%
\bibitem [{\citenamefont {Anisovich}\ and\ \citenamefont
  {Leutwyler}(1996)}]{Anisovich:1996tx}%
  \BibitemOpen
  \bibfield  {author} {\bibinfo {author} {\bibfnamefont {A.~V.}\ \bibnamefont
  {Anisovich}}\ and\ \bibinfo {author} {\bibfnamefont {H.}~\bibnamefont
  {Leutwyler}},\ }\href {https://doi.org/10.1016/0370-2693(96)00192-X}
  {\bibfield  {journal} {\bibinfo  {journal} {Phys. Lett.}\ }\textbf {\bibinfo
  {volume} {B375}},\ \bibinfo {pages} {335} (\bibinfo {year} {1996})},\ \Eprint
  {https://arxiv.org/abs/hep-ph/9601237} {arXiv:hep-ph/9601237
  [hep-ph]}\BibitemShut {NoStop}%
%%CITATION = HEP-PH/9601237;%%
\bibitem [{\citenamefont {Descotes-Genon}\ and\ \citenamefont
  {Moussallam}(2014)}]{Descotes-Genon:2014tla}%
  \BibitemOpen
  \bibfield  {author} {\bibinfo {author} {\bibfnamefont {S.}~\bibnamefont
  {Descotes-Genon}}\ and\ \bibinfo {author} {\bibfnamefont {B.}~\bibnamefont
  {Moussallam}},\ }\href {https://doi.org/10.1140/epjc/s10052-014-2946-8}
  {\bibfield  {journal} {\bibinfo  {journal} {Eur. Phys. J.}\ }\textbf
  {\bibinfo {volume} {C74}},\ \bibinfo {pages} {2946} (\bibinfo {year}
  {2014})},\ \Eprint {https://arxiv.org/abs/1404.0251} {arXiv:1404.0251
  [hep-ph]}\BibitemShut {NoStop}%
%%CITATION = ARXIV:1404.0251;%%
\bibitem [{\citenamefont {Schneider}\ \emph {et~al.}(2012)\citenamefont
  {Schneider}, \citenamefont {Kubis},\ and\ \citenamefont
  {Niecknig}}]{Schneider:2012ez}%
  \BibitemOpen
  \bibfield  {author} {\bibinfo {author} {\bibfnamefont {S.~P.}\ \bibnamefont
  {Schneider}}, \bibinfo {author} {\bibfnamefont {B.}~\bibnamefont {Kubis}},\
  and\ \bibinfo {author} {\bibfnamefont {F.}~\bibnamefont {Niecknig}},\ }\href
  {https://doi.org/10.1103/PhysRevD.86.054013} {\bibfield  {journal} {\bibinfo
  {journal} {Phys. Rev. D}\ }\textbf {\bibinfo {volume} {86}},\ \bibinfo
  {pages} {054013} (\bibinfo {year} {2012})},\ \Eprint
  {https://arxiv.org/abs/1206.3098} {arXiv:1206.3098 [hep-ph]}\BibitemShut
  {NoStop}%
\bibitem [{\citenamefont {Koepp}(1974)}]{Koepp:1974da}%
  \BibitemOpen
  \bibfield  {author} {\bibinfo {author} {\bibfnamefont {G.}~\bibnamefont
  {Koepp}},\ }\href {https://doi.org/10.1103/PhysRevD.10.932} {\bibfield
  {journal} {\bibinfo  {journal} {Phys. Rev. D}\ }\textbf {\bibinfo {volume}
  {10}},\ \bibinfo {pages} {932} (\bibinfo {year} {1974})}\BibitemShut
  {NoStop}%
\bibitem [{\citenamefont {Terschl{\"u}sen}\ \emph {et~al.}(2013)\citenamefont
  {Terschl{\"u}sen}, \citenamefont {Strandberg}, \citenamefont {Leupold},\ and\
  \citenamefont {Eichst{\"a}dt}}]{Terschlusen:2013iqa}%
  \BibitemOpen
  \bibfield  {author} {\bibinfo {author} {\bibfnamefont {C.}~\bibnamefont
  {Terschl{\"u}sen}}, \bibinfo {author} {\bibfnamefont {B.}~\bibnamefont
  {Strandberg}}, \bibinfo {author} {\bibfnamefont {S.}~\bibnamefont
  {Leupold}},\ and\ \bibinfo {author} {\bibfnamefont {F.}~\bibnamefont
  {Eichst{\"a}dt}},\ }\href {https://doi.org/10.1140/epja/i2013-13116-6}
  {\bibfield  {journal} {\bibinfo  {journal} {Eur. Phys. J. A}\ }\textbf
  {\bibinfo {volume} {49}},\ \bibinfo {pages} {116} (\bibinfo {year} {2013})},\
  \Eprint {https://arxiv.org/abs/1305.1181} {arXiv:1305.1181
  [hep-ph]}\BibitemShut {NoStop}%
\bibitem [{\citenamefont {Press}\ \emph {et~al.}(2007)\citenamefont {Press},
  \citenamefont {Teukolsky}, \citenamefont {Vetterling},\ and\ \citenamefont
  {Flannery}}]{recipes}%
  \BibitemOpen
  \bibfield  {author} {\bibinfo {author} {\bibfnamefont {W.~H.}\ \bibnamefont
  {Press}}, \bibinfo {author} {\bibfnamefont {S.~A.}\ \bibnamefont
  {Teukolsky}}, \bibinfo {author} {\bibfnamefont {W.~T.}\ \bibnamefont
  {Vetterling}},\ and\ \bibinfo {author} {\bibfnamefont {B.~P.}\ \bibnamefont
  {Flannery}},\ }\href@noop {} {\emph {\bibinfo {title} {Numerical Recipes 3rd
  Edition: The Art of Scientific Computing}}},\ \bibinfo {edition} {3rd}\ ed.\
  (\bibinfo  {publisher} {Cambridge University Press},\ \bibinfo {address} {New
  York, NY, USA},\ \bibinfo {year} {2007})\BibitemShut {NoStop}%
\bibitem [{\citenamefont {Efron}\ and\ \citenamefont
  {Tibshirani}(1994)}]{EfroTibs93}%
  \BibitemOpen
  \bibfield  {author} {\bibinfo {author} {\bibfnamefont {B.}~\bibnamefont
  {Efron}}\ and\ \bibinfo {author} {\bibfnamefont {R.}~\bibnamefont
  {Tibshirani}},\ }\href
  {https://www.crcpress.com/An-Introduction-to-the-Bootstrap/Efron-Tibshirani/p/book/9780412042317}
  {\emph {\bibinfo {title} {An Introduction to the Bootstrap}}},\ Chapman \&
  Hall/CRC Monographs on Statistics \& Applied Probability\ (\bibinfo
  {publisher} {Taylor \& Francis},\ \bibinfo {year} {1994})\BibitemShut
  {NoStop}%
\bibitem [{\citenamefont {Landay}\ \emph {et~al.}(2017)\citenamefont {Landay},
  \citenamefont {D\"oring}, \citenamefont {Fern\'andez-Ram\'irez},
  \citenamefont {Hu},\ and\ \citenamefont {Molina}}]{Landay:2016cjw}%
  \BibitemOpen
  \bibfield  {author} {\bibinfo {author} {\bibfnamefont {J.}~\bibnamefont
  {Landay}}, \bibinfo {author} {\bibfnamefont {M.}~\bibnamefont {D\"oring}},
  \bibinfo {author} {\bibfnamefont {C.}~\bibnamefont {Fern\'andez-Ram\'irez}},
  \bibinfo {author} {\bibfnamefont {B.}~\bibnamefont {Hu}},\ and\ \bibinfo
  {author} {\bibfnamefont {R.}~\bibnamefont {Molina}},\ }\href
  {https://doi.org/10.1103/PhysRevC.95.015203} {\bibfield  {journal} {\bibinfo
  {journal} {Phys.Rev.}\ }\textbf {\bibinfo {volume} {C95}},\ \bibinfo {pages}
  {015203} (\bibinfo {year} {2017})},\ \Eprint
  {https://arxiv.org/abs/1610.07547} {arXiv:1610.07547 [nucl-th]}\BibitemShut
  {NoStop}%
%%CITATION = ARXIV:1610.07547;%%
\bibitem [{\citenamefont {Terschl{\"u}sen}\ \emph {et~al.}(2012)\citenamefont
  {Terschl{\"u}sen}, \citenamefont {Leupold},\ and\ \citenamefont
  {Lutz}}]{Terschlusen:2012xw}%
  \BibitemOpen
  \bibfield  {author} {\bibinfo {author} {\bibfnamefont {C.}~\bibnamefont
  {Terschl{\"u}sen}}, \bibinfo {author} {\bibfnamefont {S.}~\bibnamefont
  {Leupold}},\ and\ \bibinfo {author} {\bibfnamefont {M.}~\bibnamefont
  {Lutz}},\ }\href {https://doi.org/10.1140/epja/i2012-12190-6} {\bibfield
  {journal} {\bibinfo  {journal} {Eur. Phys. J. A}\ }\textbf {\bibinfo {volume}
  {48}},\ \bibinfo {pages} {190} (\bibinfo {year} {2012})},\ \Eprint
  {https://arxiv.org/abs/1204.4125} {arXiv:1204.4125 [hep-ph]}\BibitemShut
  {NoStop}%
\bibitem [{\citenamefont {Ananthanarayan}\ \emph {et~al.}(2014)\citenamefont
  {Ananthanarayan}, \citenamefont {Caprini},\ and\ \citenamefont
  {Kubis}}]{Ananthanarayan:2014pta}%
  \BibitemOpen
  \bibfield  {author} {\bibinfo {author} {\bibfnamefont {B.}~\bibnamefont
  {Ananthanarayan}}, \bibinfo {author} {\bibfnamefont {I.}~\bibnamefont
  {Caprini}},\ and\ \bibinfo {author} {\bibfnamefont {B.}~\bibnamefont
  {Kubis}},\ }\href {https://doi.org/10.1140/epjc/s10052-014-3209-4} {\bibfield
   {journal} {\bibinfo  {journal} {Eur. Phys. J. C}\ }\textbf {\bibinfo
  {volume} {74}},\ \bibinfo {pages} {3209} (\bibinfo {year} {2014})},\ \Eprint
  {https://arxiv.org/abs/1410.6276} {arXiv:1410.6276 [hep-ph]}\BibitemShut
  {NoStop}%
\bibitem [{\citenamefont {Caprini}(2015)}]{Caprini:2015wja}%
  \BibitemOpen
  \bibfield  {author} {\bibinfo {author} {\bibfnamefont {I.}~\bibnamefont
  {Caprini}},\ }\href {https://doi.org/10.1103/PhysRevD.92.014014} {\bibfield
  {journal} {\bibinfo  {journal} {Phys. Rev. D}\ }\textbf {\bibinfo {volume}
  {92}},\ \bibinfo {pages} {014014} (\bibinfo {year} {2015})},\ \Eprint
  {https://arxiv.org/abs/1505.05282} {arXiv:1505.05282 [hep-ph]}\BibitemShut
  {NoStop}%
\bibitem [{\citenamefont {Oller}(2019)}]{Oller:2019opk}%
  \BibitemOpen
  \bibfield  {author} {\bibinfo {author} {\bibfnamefont {J.~A.}\ \bibnamefont
  {Oller}},\ }\href@noop {} {\  (\bibinfo {year} {2019})},\ \Eprint
  {https://arxiv.org/abs/1909.00370} {arXiv:1909.00370 [hep-ph]}\BibitemShut
  {NoStop}%
%%CITATION = ARXIV:1909.00370;%%
\bibitem [{\citenamefont {Oller}(2020)}]{Oller:2020guq}%
  \BibitemOpen
  \bibfield  {author} {\bibinfo {author} {\bibfnamefont {J.}~\bibnamefont
  {Oller}},\ }\href@noop {} {\  (\bibinfo {year} {2020})},\ \Eprint
  {https://arxiv.org/abs/2005.14417} {arXiv:2005.14417 [hep-ph]}\BibitemShut
  {NoStop}%
\bibitem [{\citenamefont {Akhmetshin}\ \emph {et~al.}(2003)\citenamefont
  {Akhmetshin} \emph {et~al.}}]{Akhmetshin:2003ag}%
  \BibitemOpen
  \bibfield  {author} {\bibinfo {author} {\bibfnamefont {R.}~\bibnamefont
  {Akhmetshin}} \emph {et~al.} (\bibinfo {collaboration} {CMD-2}),\ }\href
  {https://doi.org/10.1016/S0370-2693(03)00595-1} {\bibfield  {journal}
  {\bibinfo  {journal} {Phys. Lett. B}\ }\textbf {\bibinfo {volume} {562}},\
  \bibinfo {pages} {173} (\bibinfo {year} {2003})},\ \Eprint
  {https://arxiv.org/abs/hep-ex/0304009} {arXiv:hep-ex/0304009}\BibitemShut
  {NoStop}%
\bibitem [{\citenamefont {Achasov}\ \emph {et~al.}(2012)\citenamefont {Achasov}
  \emph {et~al.}}]{Achasov:2012zza}%
  \BibitemOpen
  \bibfield  {author} {\bibinfo {author} {\bibfnamefont {M.}~\bibnamefont
  {Achasov}} \emph {et~al.},\ }\href
  {https://doi.org/10.1134/S0021364011220024} {\bibfield  {journal} {\bibinfo
  {journal} {JETP Lett.}\ }\textbf {\bibinfo {volume} {94}},\ \bibinfo {pages}
  {734} (\bibinfo {year} {2012})}\BibitemShut {NoStop}%
\bibitem [{\citenamefont {Achasov}\ \emph {et~al.}(2013)\citenamefont {Achasov}
  \emph {et~al.}}]{Achasov:2013btb}%
  \BibitemOpen
  \bibfield  {author} {\bibinfo {author} {\bibfnamefont {M.}~\bibnamefont
  {Achasov}} \emph {et~al.},\ }\href
  {https://doi.org/10.1103/PhysRevD.88.054013} {\bibfield  {journal} {\bibinfo
  {journal} {Phys. Rev. D}\ }\textbf {\bibinfo {volume} {88}},\ \bibinfo
  {pages} {054013} (\bibinfo {year} {2013})},\ \Eprint
  {https://arxiv.org/abs/1303.5198} {arXiv:1303.5198 [hep-ex]}\BibitemShut
  {NoStop}%
\bibitem [{\citenamefont {Achasov}\ \emph {et~al.}(2016)\citenamefont {Achasov}
  \emph {et~al.}}]{Achasov:2016zvn}%
  \BibitemOpen
  \bibfield  {author} {\bibinfo {author} {\bibfnamefont {M.}~\bibnamefont
  {Achasov}} \emph {et~al.},\ }\href
  {https://doi.org/10.1103/PhysRevD.94.112001} {\bibfield  {journal} {\bibinfo
  {journal} {Phys. Rev. D}\ }\textbf {\bibinfo {volume} {94}},\ \bibinfo
  {pages} {112001} (\bibinfo {year} {2016})},\ \Eprint
  {https://arxiv.org/abs/1610.00235} {arXiv:1610.00235 [hep-ex]}\BibitemShut
  {NoStop}%
\end{thebibliography}%

%***********************************************
\end{document}